\documentclass[pra,10pt,twocolumn,superscriptaddress]{revtex4-2}

\usepackage{graphicx}
\usepackage{dcolumn}
\usepackage{bm}
\usepackage{amsmath}
\usepackage[thinc]{esdiff}
\usepackage{physics}
\usepackage{tikz}
\usepackage{tikz-qtree}
\usetikzlibrary{trees}
\usepackage{float}
\usepackage{dsfont}
\usepackage{amssymb}
\usepackage[section]{placeins}
\usepackage{hyperref}

\usepackage{epsfig}
\usepackage{amsmath}
\usepackage{graphicx}
\usepackage{graphics}
\usepackage{float}
\usepackage{amssymb}
\usepackage{natbib}
\usepackage{epstopdf}
\usepackage{verbatim}

\hypersetup{
    colorlinks = true,
    allcolors = blue
}

\begin{document}

\preprint{APS/123-QED}

\title{Continuous dynamical decoupling of spin chains: Inducing two-qubit interactions to generate perfect entanglement}

\author{Abdullah Irfan}
\author{Syed Furqan Abbas Hashmi}
\author{Syeda Neha Zaidi}
\author{Muhammad Usman Baig}
\author{Wahaj Ayub}
\author{Adam Zaman Chaudhry}
 \email{adam.zaman@lums.edu.pk}
\affiliation{School of Science and Engineering, Lahore University of Management Sciences (LUMS), Opposite Sector U, D.H.A, Lahore 54792, Pakistan}%




\begin{abstract}
Efficient control over entanglement in spin chains is useful for quantum information processing applications. In this paper, we propose the use of a combination of two different configurations of strong static and oscillating fields to control and generate near-perfect entanglement between any two spins in a spin chain, even in the presence of noise. This is made possible by the fact that our control fields not only decouple the spin chain from its environment but also selectively modify the spin-spin interactions. By suitably tuning these spin-spin interactions via the control fields, we show that the quantum state of any two spins in the spin chain can be made to be a Bell state. We illustrate our results for various spin chains, such as the XY model, the XYZ model, and the Ising spin chain.
\end{abstract}


\maketitle


\section{\label{sec:intro}Introduction}
Quantum spin chains have been the subject of extensive theoretical and experimental studies \cite{c1}. They are paradigmatic systems providing convenient and tractable models that yield insights into a range of physical phenomena \cite{c2,c3,c4}. In particular, spin chains have attracted considerable attention due to their potential use in quantum information applications such as quantum computation and short-distance quantum communication \cite{c1}. Achieving entanglement between spins is imperative when it comes to quantum information applications, and has been a long-standing  goal  and  a  focus  of  many  studies \cite{SpillerNJP2007,LeonardoPRL2011,HosseinPRA2013,SahlingNatPhys2015,EstarellasPRA2017}. For instance, long-range Ising-type interactions \cite{c10}, local effective magnetic fields~\cite{c11}, staggered magnetic fields \cite{c12}, and dynamical decoupling~\cite{c6} have been made to realize this goal. Driven spin chains as high-quality quantum routers, which generate highly entangled states over arbitrary distances in spin chains by applying external fields, have also been proposed \cite{c38}. Spin chains may also act as quantum mediators in order to achieve perfect long-range entanglement between remote spins in bulk and on the surfaces of magnetic nanostructures~\cite{c7}. 

Like any physical system, spin chains interact with their environment, which eventually leads to the loss of the quantum character of the system, a phenomenon known as decoherence \cite{c13}. Moreover, entangled states are very fragile when exposed to the environment \cite{c14}.  This poses a serious problem when it comes to using the quantum state of the spin chain for quantum information processing tasks \cite{c15,c16,c18,c19}.
One reliable method of suppressing the system-environment interaction of an open quantum system is dynamical decoupling \cite{c20,c21,c22,c23,c24,c25,c26,c27,c28,c29,c30}. Dynamical decoupling works by averaging out unwanted effects of the environment interaction through the application of control fields which effectively modulate the system-environment interaction \cite{c6}. In the context of spin chains, Ref.~\cite{c6} has shown that the application of strong static and oscillating fields not only protects the spin chain from decoherence, but also modulates the spin-chain interaction such that the effective Hamiltonian of the spin chain includes interaction terms that are not present in the original spin chain Hamiltonian. This shows that in addition to protecting the spin chain from the environment, control fields can potentially be used to achieve perfect quantum state transfer and improved two-spin entanglement generation in the spin chain. 

Our aim in this paper is to look for a configuration of control fields that selectively suppresses the spin-spin interactions in addition to decoupling the chain from the environment. If such control fields exist, can we devise a scheme that allows us to perfectly entangle spins in the chain while protecting it from decoherence? We answer this question by showing that we can apply different control fields to the even and odd indexed sites in a chain in order to decouple the spins from the environment and remove spin-spin interactions in the chain, effectively obtaining a chain of isolated spins or qubits. We refer to this configuration of fields as a staggered configuration. This then opens up the possibility of inducing two-qubit interactions in the spin chain by applying the same control fields to two spins and staggered fields to the rest of the spin chain. We propose a scheme that allows us to use such two-qubit interactions to entangle any two spins in the spin chain. We note that simply decoupling the spin chain from the environment using control fields may also result in a modulated spin chain Hamiltonian that generates entanglement in the chain~\cite{c6}. Such a modulated Hamiltonian may be considered to be a consequence of decoupling the spin chain from the environment, and therefore entangles different spins to varying degrees depending on the their positions and on the original spin chain Hamiltonian. Generally speaking, the entanglement generation is far from ideal. The technique we discuss in this paper generates entanglement by first suppressing spin-spin interactions in the chain, and then using a sequence of two-qubit interactions to entangle any two spins of our choice. As a result, we have greater control over the entanglement process, allowing us to predict the duration of each interaction, and the state of the spin chain when it is perfectly entangled. 

We begin by considering a general one-dimensional spin chain that interacts with its environment \cite{c32}. We show that a staggered configuration of control fields can not only protect the spin chain from the environment, but it can also suppress spin-spin interactions in the chain. This is achieved by proving that the time-averaged effective Hamiltonian is zero when staggered control fields are applied. Applying the same control fields to spins is referred to as a constant configuration. By applying a constant configuration of control fields to two neighboring spins, and a staggered configuration of control fields to the rest of the spin chain, we can induce a two-qubit interaction that can perfectly entangle the two neighboring spins. We then propose a scheme that uses a sequence of two-qubit interactions between neighboring spins to perfectly entangle any two spins in the chain. As an example, we solve this scheme for the XY chain, showing analytically that any two spins in the chain can be perfectly entangled by choosing appropriate interaction durations. Since the interaction is only between two spins at any point in time, we essentially solve a system of two spins evolving under the action of the effective Hamiltonian. We find a way to remarkably simplify the quantum state of the spin chain at the end of each two-qubit interaction - the generation of entanglement is then evident. Thereafter, we apply our scheme to other spin chains such as the XYZ model and present numerical results conclusively demonstrating the effectiveness of our scheme.

This paper is organized as follows. In Sec.~\ref{sec:formalism} we show how a staggered configuration of static and oscillating control fields can be used to decouple a spin chain from the environment and also suppress spin-spin interactions in the chain. Sec.~\ref{sec:scheme} considers the XY model and describes how a combination of constant and staggered control fields can be used to induce two-qubit interactions, which can be used to generate perfect entanglement in the spin chain. Sec.~\ref{sec:numerical} provides results from numerical simulations that corroborate our presented scheme, and shows how our proposed method can be used for other spin chains such as the Ising chain and the XYZ model. We conclude the paper in Sec.~\ref{sec:conclusion}.

\section{\label{sec:formalism}Formalism}
We begin by showing how strong static and oscillating control fields can be used to decouple a spin chain from its environment. Instead of using the same control fields for every site in the spin chain, we propose a configuration of control fields such that there is one type of control field for spins at odd numbered sites, and another type of control field for spins at even numbered sites. We call this a staggered configuration. This achieves a two-fold task. First, it decouples the spin chain from its environment to lowest order. Second, it suppresses the spin-spin interactions in the chain, effectively producing a chain of non-interacting qubits. Note that if the same field is applied to all the spins, the spin-spin interactions are not removed \cite{c6}. 

Consider then a spin chain with nearest neighbor interactions. We write the Hamiltonian as (we take $\hbar = 1$ throughout) 
\begin{equation}
H_{0} = \sum_{j = 1}^{N-1} \sum_{k=1}^{3} \zeta_{jk}\sigma_{k}^{(j)}\sigma_{k}^{(j+1)},
\label{eq:H0}
\end{equation} 
where $\zeta_{jk}$ are the coupling strengths between the spins, $j$ labels the sites, and $k = 1,2,3 $ denotes $x$, $y$, and $z$, respectively. The Pauli spin operators follow the usual commutation relations $[\sigma_{j}^{(p)},\sigma_{k}^{(m)}] = 2 i \delta_{pm} \epsilon_{jkl} \sigma_{l}^{(p)}$. Note that we are not using cyclic boundary conditions.
We assume the spin chain interaction with its environment to be given by 
\begin{equation}
H_{\text{SB}} =  \sum_{j = 1}^{N-1} \sum_{k=1}^{3} B_{k}^{(j)}\sigma_{k}^{(j)}, 
\label{eq:HSB}
\end{equation}
where $B_{k}^{(j)}$ are arbitrary environment operators; they can also represent randomly fluctuating noise terms for a classical bath. Our first task is to find periodic control fields that decouple the spin chain from its environment, at least to lowest order. Corresponding to the control fields is a unitary operator $U_{c}(t)$ that satisfies $i \diffp{U_c(t)}{t} = H_{c}(t) U_{c}(t)$, where $H_{c}(t)$ is the Hamiltonian that describes the action of the control fields on the system. Moreover, the fields we consider for this task are periodic; the unitary operator satisfies $U_{c}(t+t_c) = U_{c}(t)$, where $t_c$ is the time period. In order for the control fields to decouple the spin chain from the environment to lowest order, we must have that \cite{c31,c32,c33,c34}
\begin{equation}
\int_{0}^{t_c} dt \, U_{c}^{\dagger}(t)H_{\text{SB}}U_{c}(t)=0
\label{eq:decoupling}
\end{equation}
Keeping in mind the form of the interaction between the spin chain and the environment, we guess that a field configuration represented by the following unitary operator may decouple the spin chain from its environment:
\begin{eqnarray}
U_{c}(t)=&\Biggl(\prod_{i = 1,3,5,...}^{N}e^{i\omega n_{x} \sigma_{x}^{(i)}t}e^{i\omega n_{y} \sigma_{y}^{(i)}t}\Biggr)\nonumber \\
&\times \Biggl(\prod_{i = 2,4,6,...}^{N-1}e^{i\omega m_{x} \sigma_{x}^{(i)}t}e^{i\omega m_{y} \sigma_{y}^{(i)}t}\Biggr).
\end{eqnarray}
For concreteness, here we have considered $N$ to be odd; the case of even $N$ is dealt with in a similar fashion. Here $\omega = 2\pi/t_c$, and the integers $n_x$, $n_y$ (for spins at odd indexed sites) and $m_x$, $m_y$ (for spins at even indexed sites) differentiate between the two types of control fields applied to alternate spins. In other words, $U_c(t)$ represents a staggered field configuration in which different control fields are applied to spins located at odd and even-numbered sites in the chain. Now, it is obvious that our unitary operator $U_c(t)$ satisfies $U_{c}(t+t_c) = U_{c}(t)$. Our task then is to show that this configuration of control fields satisfies the decoupling condition given in Eq.~\eqref{eq:decoupling}. We expect this to be the case - loosely put, $e^{i\omega n_{x} \sigma_{x}^{(i)}t}$ averages out the contribution of the noise due to $\sigma_y^{(i)}$ and $\sigma_z^{(i)}$, while $e^{i\omega n_{y} \sigma_{y}^{(i)}t}$ takes care of the remaining noise. To verify that this is indeed the case, it is useful to define
\begin{equation*}
h_{j,k,p}(t) = U_{c}^{\dagger}(t)\sigma_{k}^{(j)}U_{c}(t),
\end{equation*}
where $k=1,2,3$, $\sigma^{(j)}_{1}=\sigma^{(j)}_{x}$, $\sigma^{(j)}_{2}=\sigma^{(j)}_{y}$, $\sigma^{(j)}_{3}=\sigma^{(j)}_{z}$, and $p = j \, \text{mod} \, 2$. The index $p$ tells us if the constants used in the control field at site $j$ are $n_x, n_y$ or $m_x, m_y$; $p = 0$ means that the field represented by $n_x, n_y$ is being applied, and $p = 1$ means that the field represented by $m_x, m_y$ is being applied.  Using the commutation relations for the Pauli spin operators, it is straightforward to show that 
\begin{align*}
h_{j,1, 0}(t) =& \cos(2\omega n_{y} t ) \sigma_{x}^{(j)} - \sin(2\omega n_{y} t ) \sigma_{z}^{(j)},\\
h_{j,2, 0}(t) =& \sin(2\omega n_{x}t )\sin(2 \omega n_{y} t)\sigma_{x}^{(j)} + \cos(2 \omega n_{x}t) \sigma_{y}^{(j)} 
\\ &+\sin(2 \omega n_{x}t) \cos(2 \omega n_{y} t )\sigma_{z}^{(j)},\\
h_{j,3, 0}(t) =& \cos(2\omega n_{x}t)\sin(2\omega n_{y}t)\sigma_{x}^{(j)} - \sin(2\omega n_{x}t)\sigma_{y}^{(j)} 
\\ &+ \cos(2 \omega n_{x}t)\cos(2 \omega n_{y}t)\sigma_{z}^{(j)}, \\ 
\end{align*}
$h_{j,1,1}(t)$, $h_{j,2,1}(t)$, and $h_{j,3,1}(t)$ are the same as $h_{j,1,0}(t)$, $h_{j,2,0}(t)$, and $h_{j,3,0}(t)$ respectively, except that $n_x$ is replaced by $m_x$ and $n_y$ by $m_y$. With these relations, it is easy to show that Eq.~\eqref{eq:decoupling} is satisfied if $n_x \neq n_y$ and $m_x \neq m_y$, meaning that $U_c(t)$ effectively decouples the spin chain from its environment (at least to lowest order). The control field Hamiltonian corresponding to $U_c(t)$ is found from the Schrodinger equation to be  
\begin{align}
H_{c}(t) =& \sum_{i = 1,3,5,...}^{N} \{ \omega n_y  \bigl[ \sin(2 \omega n_x t)\sigma_{z}^{(i)} \nonumber \\
 -& \cos(2 \omega n_x t)\sigma_{y}^{(i)} \bigr] - \omega n_x \sigma_{x}^{(i)}  \} \nonumber \\
 +& \sum_{i = 2,4,8,...}^{N} \{ \omega m_y  \bigl[ \sin(2 \omega m_x t)\sigma_{z}^{(i)} \nonumber \\
 -& \cos(2 \omega m_x t)\sigma_{y}^{(i)} \bigr] - \omega m_x \sigma_{x}^{(i)}  \},
\label{eq:Hc}
\end{align}
with $n_x \neq n_y$ and $m_x \neq m_y$. We must point out that the spin chain is decoupled from the environment only if the fields oscillate fast enough. More precisely, decoupling occurs if $ t_c \ll \tau_c$ where $\tau_c$ is the environment correlation time \cite{KurizkiPRA2011}. 

In addition to decoupling the spin chain from its environment, our staggered control field configuration can also suppress the spin-spin interactions in the chain. To show this, we must find the effective Hamiltonian of the spin chain when control fields are applied. It is known that if the control fields are strong enough and oscillate fast enough (that is, faster than the typical timescale of the evolution due to the spin chain Hamiltonian itself), the effective Hamiltonian can be written as~\cite{c32,c33}
\begin{equation}
\Bar{H} = \frac{1}{t_{c}}\int_{0}^{t_{c}} dt \, U_{c}^{\dagger}(t)H_{0}U_{c}(t),
\end{equation}
where $H_0$ is the original spin chain Hamiltonian given in Eq.~\eqref{eq:H0}. For our case, the effective Hamiltonian simplifies to 
\begin{equation}
\Bar{H} = \frac{1}{t_{c}}\sum_{j=1}^{N-1}\int_{0}^{tc}dt \sum_{k=1}^{3}\zeta_{jk}h_{j,k,p}(t)h_{j+1,k,p'}(t).
\label{eq:Heff}
\end{equation}
Note that $p' = (j+1)\, \text{mod} \, 2$.
Because one of the $j$ and $j+1$ sites is odd while the other is even, one of $p$ and $p'$ must be equal to one while the other must be zero. To simplify Eq.~\eqref{eq:Heff} further, we define the operators
\begin{equation}
I_{k}^{(j)} = \frac{1}{t_{c}}\int_{0}^{t_{c}}h_{j,k,p}(t)h_{j+1,k,p'}(t) dt.
\end{equation}
These allow us to write the effective Hamiltonian succinctly as
\begin{equation}
\Bar{H} = \sum_{j=1}^{N-1}\sum_{k=1}^{3}\zeta_{jk}I_{k}^{(j)}.
\label{eq:effH}
\end{equation}
To explicitly obtain an expression for $\Bar{H}$, we must evaluate the integrals $I_{k}^{(j)}$. To remove the spin-spin interactions in the spin chain, at least to lowest order, one possible choice, which we stick to in this paper, is to choose $n_y = 2 n_x$, $m_y = 2m_x
$, and $n_x \neq m_x$. The integrals in $I_{k}^{(j)}$ then evaluate to zero for all $j$ and $k$, leading to $\Bar{H} = 0$. Notice that since $n_x \neq m_x$, the suppression of the spin-spin interactions depends crucially on the fact that we apply different control fields to odd and even-numbered sites. On the other hand, let us examine what happens if we apply the same fields to two neighboring spins in the chain. In other words, we have a `constant' configuration of the control fields for the two spins. In this case, we find that 
\begin{align}
I_{1}^{(j)} &= \frac{1}{2} \bigl(\sigma_{x}^{(j)} \sigma_{x}^{(j+1)} + \sigma_{z}^{(j)} \sigma_{z}^{(j+1)} \bigr), \label{eff1}  \\
I_{2}^{(j)} &= \frac{1}{4} \bigl(\sigma_{x}^{(j)} \sigma_{x}^{(j+1)} + 2\sigma_{y}^{(j)} \sigma_{y}^{(j+1)} + \sigma_{x}^{(j)} \sigma_{y}^{(j+1)} \notag \\
&+ \sigma_{y}^{(j)} \sigma_{x}^{(j+1)} + \sigma_{z}^{(j)} \sigma_{z}^{(j+1)} \bigr) , \label{eff2}  \\
I_{3}^{(j)} &= \frac{1}{4} \bigl( \sigma_{x}^{(j)} \sigma_{x}^{(j+1)} + 2\sigma_{y}^{(j)} \sigma_{y}^{(j+1)} - \sigma_{x}^{(j)} \sigma_{y}^{(j+1)} \notag \\
&- \sigma_{y}^{(j)} \sigma_{x}^{(j+1)} + \sigma_{z}^{(j)} \sigma_{z}^{(j+1)} \bigr)\label{eff3}.
\end{align}
The effective Hamiltonian is now not zero. Moreover, since $n_y \neq n_x$, the two spins are still decoupled from their environment. To sum up then, both the constant and staggered configurations of control fields can decouple the spin chain form its environment. While the staggered configuration suppresses spin-spin interactions, the constant configuration modulates the spin-spin interactions, giving us an effective Hamiltonian that describes the interaction of the spins. We now show how we can use both both staggered and constant field configurations in conjunction to generate entanglement between specific spins in the chain. Note that the effect of the environment is also suppressed.

\section{\label{sec:scheme}Scheme for entanglement generation}

For concreteness, we focus on generating perfect entanglement between the first spin and an arbitrary spin in the spin chain. Our considerations can be generalized in a very straightforward manner to any two spins in the chain. We start by noting that the expression for $I_{k}^{(j)}$  contains terms from two neighboring spins, namely $h_{j,k,p}$ and $h_{j+1,k,p'}$ where $j$ is the site of the spin. Whether or not $I_{k}^{(j)}$ and subsequently the effective Hamiltonian for these two neighboring spins is zero depends on whether the integers defining the control field at both sites are equal or not. Keeping this fact in mind, we propose a field configuration such that for the first two spins of the chain, the integers defining the control field are equal, and for the rest of the chain, no two neighboring spins have a control field defined by the same integers. In other words, we have a constant configuration for the first two spins and a staggered configuration for the rest of the chain. This allows the first two spins to interact while the rest of the spin chain is effectively dormant and non-interacting. The effective Hamiltonian can be written as 
\begin{equation}
\Bar{H} = \Biggl(\sum_{k=1}^{3}\zeta_{1k}I_{k}^{(1)}\Biggr) \otimes \mathds{1}^{(3)} \otimes \mathds{1}^{(4)} \otimes ... \otimes \mathds{1}^{(N)},
\end{equation}
where $\mathds{1}$ denotes the identity operator. 
Depending on the interaction, the effective Hamiltonian $\Bar{H}$ may entangle the first two spins to some degree. For example, one can prove that if we consider the XY chain with constant coupling strengths ($\zeta_{j1} = \zeta_{j2}=1$ for convenience, $\zeta_{j3}=0$), the first two spins become perfectly entangled if the spin is left to evolve under $\Bar{H}$ for a certain duration. We label this duration $\tau_1$. We then make a change. We now set the integers for the second and third spins in the chain equal while keeping the integers of neighboring sites different for all other neighboring spins in the chain. Following our previous reasoning, it is clear that now the second and third spins interact, while the remaining spins are non-interacting. The effective Hamiltonian is now
\begin{equation}
\Bar{H} = \mathds{1}^{(1)} \otimes \Biggl(\sum_{k=1}^{3}\zeta_{2k}I_{k}^{(2)}\Biggr) \otimes \mathds{1}^{(4)}  \otimes ... \otimes \mathds{1}^{(N)}.
\end{equation}
The idea is to now allow the spin chain to evolve under this effective Hamiltonian until the first and the third spins are perfectly entangled. The time required for this interaction is labeled $\tau_2$. This process can be continued until the first spin is perfectly entangled with any spin that we wish to entangle it with. 

Let us now demonstrate this scheme of generating entanglement in detail for the XY spin chain with $N$ sites. We set $\zeta_{j1} = \zeta_{j2} = 1$ and $\zeta_{j3} = 0$ in Eq.~\eqref{eq:H0}. We begin by setting the initial state of the system as
\begin{equation*}
\ket{\psi_{0}(0)}=\ket{0}_{1}\otimes\ket{0}_{2}\otimes...\otimes\ket{0}_{N}.
\end{equation*}
We define the states $\ket{0}$ and $\ket{1}$ as the eigenstates of $\sigma_z$ with eigenvalues $+1$ and $-1$ respectively. The effective Hamiltonian that describes the interaction between two neighboring spins $i$ and $i + 1$ under the action of the same control fields applied to the two spins is [see Eqs.~\eqref{eff1} and \eqref{eff2}]
\begin{align}
\Bar{H}_i = &\frac{1}{4}\bigg[3\sigma_{x}^{(i)}\sigma_{x}^{(i+1)} + 2\sigma_{y}^{(i)}\sigma_{y}^{(i+1)} + 3\sigma_{z}^{(i)}\sigma_{z}^{(i+1)} + \notag \\
&\sigma_{x}^{(i)}\sigma_{y}^{(i+1)} + \sigma_{y}^{(i)}\sigma_{x}^{(i+1)}\bigg].
\label{eq:b}
\end{align}
To find the evolution generated by this two-qubit interaction, we find that the eigenstates of this effective Hamiltonian, written in terms of the basis states $\ket{00}_{i,i+1}$, $\ket{01}_{i,i+1}$, $\ket{10}_{i,i+1}$, and $\ket{11}_{i,i+1}$ are 
\begin{gather*}
    \ket{e_1} = 
    \begin{bmatrix}
    0  \\ -\beta \\ \beta\\ 0
    \end{bmatrix},\,\,
    \ket{e_2} = 
    \begin{bmatrix}
    \alpha^{\ast} \\ 0 \\ 0 \\ -\beta
    \end{bmatrix},\,\,
    \ket{e_3} = 
    \begin{bmatrix}
    0 \\ \; \beta \; \\ \;\;\beta \;\; \\ 0
    \end{bmatrix},\,\,
    \ket{e_4} =
    \begin{bmatrix}
    \alpha^{\ast} \\ 0 \\ 0 \\ \beta
    \end{bmatrix},
\end{gather*}
where $\alpha=(1+2i)/\sqrt{10}$ and $\beta=1/\sqrt{2}$.
The corresponding eigenvalues are $\lambda_{1}=-2, \lambda_{2}=(3-\sqrt{5})/4, \lambda_{3}=1/2 \text{ and } \lambda_{4}=(3+\sqrt{5})/4$.
Consequently,
\begin{align*}
\ket{00}_{i,i+1} &= \alpha \ket{e_2} + \alpha \ket{e_4} \\
\ket{01}_{i,i+1} &= -\beta \ket{e_1} + \beta \ket{e_3} \\ 
\ket{10}_{i,i+1} &= \beta \ket{e_1} + \beta \ket{e_3} \\
\ket{11}_{i,i+1} &= -\beta \ket{e_2} + \beta \ket{e_4}.
\end{align*}
We now return to our original problem. With the first two spins interacting, the state at time $t$ is 
\begin{equation}
\ket{\psi(t)} = \left(\gamma_{1}(t) \ket{00}_{1,2} + \gamma_{2}(t) \ket{11}_{1,2}\right)\ket{\mathbf{0}},
\end{equation}
where 
\begin{align}
\gamma_{1}(t) &= \abs{\alpha}^2 \Bigl( e^{-i \lambda_4 t} + e^{-i \lambda_2 t} \Bigr), \label{eq:gamma1} \\
\gamma_{2}(t) &= \alpha \beta \Bigl( e^{-i \lambda_4 t} - e^{-i \lambda_2 t} \Bigr), \label{eq:gamma2}
\end{align}
and $\ket{\mathbf{0}}$ denotes that the other spins are all in the state $\ket{0}$. This evolution is consistent with the fact that $\sigma_z^{(i)}\sigma_z^{(i + 1)}$ commutes with the effective Hamiltonian. The idea now is to continue this evolution until time $\tau_1$, which is the time required for spins $1$ and $2$ to become perfectly entangled. That this is indeed possible can easily be shown by calculating the concurrence \cite{c40}; this can also be seen in the results that we present in the next section. The next step is to make spins $2$ and $3$ interact. To find the subsequent evolution, let us first note that we can write
\begin{align*}
&\ket{\psi(\tau_1)} \notag \\
&= \left[\ket{0}_1 \otimes \Bigl( \gamma_{1}(\tau_1) \ket{00}_{2,3} \Bigr) + \ket{1}_1 \otimes \Bigl( \gamma_{2}(\tau_1) \ket{10}_{2,3} \Bigr)\right]\ket{\mathbf{0}}.
\end{align*}
Now,
\begin{align*}
e^{-i \Bar{H}_i t} \ket{00}_{i,i+1} &= \gamma_{1}(t) \ket{00}_{i,i+1} + \gamma_{2}(t) \ket{11}_{i,i+1},\nonumber \\
e^{-i \Bar{H}_i t} \ket{10}_{i,i+1} &= \eta_{1}(t) \ket{01}_{i,i+1} + \eta_{2}(t) \ket{10}_{i,i+1},
\end{align*}
where $\gamma_{1}(t)$ and $\gamma_{2}(t)$ are given by Eqs.~\eqref{eq:gamma1} and \eqref{eq:gamma2}, and $\eta_{1}(t)$ and $\eta_{2}(t)$ are given by
\begin{align}
\eta_{1}(t) &= -\beta^{2}\Bigl( e^{-i \lambda_1 t} - e^{-i \lambda_3 t} \Bigr), \\
\eta_{2}(t) &= \beta^{2} \Bigl( e^{-i \lambda_1 t} + e^{-i \lambda_3 t} \Bigr).
\end{align}
These results can then be used to work out the evolution of the spin chain due to the interaction between the second and third spins. Thereafter, the third spin and the fourth spin interact, and so on. The state of the spin chain becomes increasingly complicated. Noticing that $e^{-i \Bar{H}_i t}$ acts on both $\ket{00}_{i,i+1}$ and $\ket{10}_{i,i+1}$ to produce a linear combination of $\ket{00}_{i,i+1}$ and $\ket{10}_{i,i+1}$, we can represent the evolution of the spin chain by using the tree diagram shown in Fig.~\ref{fig:tree}.

\begin{widetext}

\begin{figure}
\begin{tikzpicture}[level distance=1.8in,sibling distance=.1in,scale=0.75]
\tikzset{edge from parent/.style= 
    {thick, draw,
        edge from parent fork right},every tree node/.style={minimum width=0.8in,text width=1.1in, align=center},grow'=right}
\Tree 
[. {$\ket{00}$} 
[.{$\gamma_{1}(\tau_1)\ket{000}$}
[.{$\gamma_{1}(\tau_1)\gamma_{1}(\tau_2)\ket{0000}$}
[.{$\gamma_{1}(\tau_1)\gamma_{1}(\tau_2)\gamma_{1}(\tau_3)\ket{00000}$}
]
[.{$\gamma_{1}(\tau_1)\gamma_{1}(\tau_2)\gamma_{2}(\tau_3)\ket{00110}$}
]
]
[.{$\gamma_{1}(\tau_1)\gamma_{2}(\tau_2)\ket{0110}$}
[.{$\gamma_{1}(\tau_1)\gamma_{2}(\tau_2)\eta_{1}(\tau_3)\ket{01010}$}
]
[.{$\gamma_{1}(\tau_1)\gamma_{2}(\tau_2)\eta_{2}(\tau_3)\ket{01100}$}
]
]
]
[.{$\gamma_{2}(\tau_1)\ket{110}$}
[.{$\gamma_{2}(\tau_1)\eta_{1}(\tau_2)\ket{1010}$}
[.{$\gamma_{2}(\tau_1)\eta_{1}(\tau_2)\eta_{1}(\tau_3)\ket{10010}$}
]
[.{$\gamma_{2}(\tau_1)\eta_{1}(\tau_2)\eta_{2}(\tau_3)\ket{10100}$}
]
]
[.{$\gamma_{2}(\tau_1)\eta_{2}(\tau_2)\ket{1100}$}
[.{$\gamma_{2}(\tau_1)\eta_{2}(\tau_2)\gamma_{1}(\tau_3)\ket{11000}$}
]
[.{$\gamma_{2}(\tau_1)\eta_{2}(\tau_2)\gamma_{2}(\tau_3)\ket{11110}$}
]
]
] 
]
\end{tikzpicture}
\caption{A tree diagram is a useful way to keep track of the evolution of the spin chain. Each state vector in this diagram should be appended with $\ket{\mathbf{0}}$, meaning that the other spins are in the state $\ket{0}$. As we continue evolving the spin chain using the two-qubit Hamiltonian, the state of the chain becomes a linear combination of a larger number of different state vectors. The complete state just after each interaction is given by the sum of the states in each column. For example, the sum of the four terms in the third column represents the state of the system when spins $1$ and $3$ have interacted.}
\label{fig:tree}
\end{figure}

\end{widetext}

\begin{figure}[b!]
    \centering
    \includegraphics[width=.9\columnwidth]{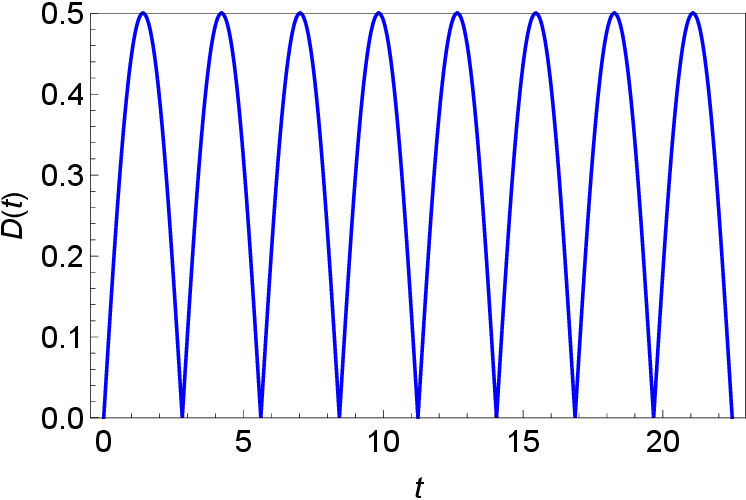}
    \caption{A plot of $D(t)$ against time, where $D(t) = \abs{\gamma_1(t)\,\gamma_2(t)}$.}
    \label{fig:det1}
\end{figure}

\begin{figure}[t!]
   \includegraphics[width=.9\columnwidth]{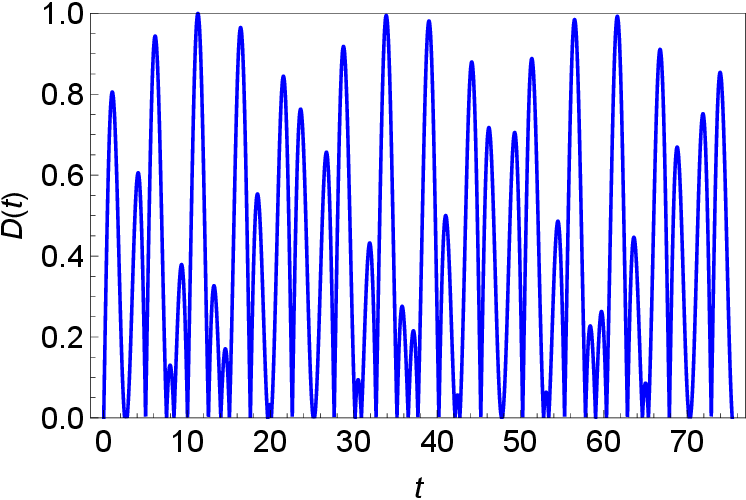}  
   \caption{A plot of $D(t)$ against time, where now $D(t) = \abs{\gamma_1(t)\,\eta_1(t)}$.}
   \label{fig:det2}
\end{figure}

We now use the worked out evolution of the spin chain to find suitable interaction durations $\tau_1, \tau_2, ...  ,\tau_i$ such that spins $1$ and $i + 1$ are perfectly entangled. We propose a question that could help us find out how this can be done: can the interaction durations be chosen such that the two-qubit state of the spins $1$ and $i + 1$ ends up in a Bell state with the rest of the spins in the $\ket{0}$ state? For this to be possible, we can see in our tree diagram that only two terms in the complete spin chain state must be non-zero. These are the terms given by 
\begin{equation*}
\left[\prod_{j=1}^{i}\gamma_{1}(\tau_j)\right] \ket{0}_1 \otimes  \prod_{j=2}^{i}\ket{0}_j \otimes \ket{0}_{i+1},
\end{equation*}
and
\begin{equation*}
\gamma_{2}(\tau_1)\left[\prod_{j=2}^{i} \eta_{1}(\tau_j)\right] \ket{1}_1 \otimes \prod_{j=2}^{i}\ket{0}_j \otimes \ket{1}_{i+1}.
\end{equation*}
If the coefficients of all the other terms in the state are zero, we are simply left with the state
\begin{align}
\ket{\psi_f} &= \Bigg[\left(\prod_{j=1}^{i}\gamma_1(\tau_{j})\right)\ket{00}_{1,i+1} + \notag \\
&\gamma_2(\tau_1) \left(\prod_{j=2}^{i}  \eta_1(\tau_{j})\right)\ket{11}_{1,i+1}\Bigg]\otimes \ket{\mathbf{0}}. \label{finalstate}
\end{align}
The two-qubit state of the first spin and the $i + 1$ spin is then obvious. This state is fully entangled if 
\begin{equation}
\abs{\gamma_1(\tau_{1})\,\gamma_2(\tau_{1})\Bigl(\prod_{j=2}^{i}\gamma_1(\tau_{j})\,\eta_1(\tau_{j})\Bigr)} = 0.5.
\label{eq:det}
\end{equation}
Notice that $\gamma_1(\tau_1)\,\gamma_2(\tau_1)$ comes from the first two-qubit interaction. Each additional two-qubit interaction is responsible for a $\gamma_1(\tau_j)\,\eta_1(\tau_j)$ term. Fig.~\ref{fig:det1} shows that $\tau_1$ can be chosen such that $\abs{\gamma_1(\tau_1)\,\gamma_2(\tau_1)} = 0.5$. Moreover, Fig.~\ref{fig:det2} shows that the interaction duration can always be chosen such that $\abs{\gamma_1(\tau_j)\,\eta_1(\tau_j)} = 1$ with $j \geq 2$. As a result, we conclude that Eq.~\eqref{eq:det} can always be satisfied by choosing suitable interaction durations. We have also checked that if we choose such interaction durations, the coefficients of all other terms in the complete state of the spin chain are zero, justifying writing the state of the spin chain in the form given by Eq.~\eqref{finalstate}. Since the first interaction duration is obtained by finding a maximum of $\abs{\gamma_1(\tau_1)\,\gamma_2(\tau_1)}$ and all other interaction durations are obtained by finding a maximum of $\abs{\gamma_1(\tau_j)\,\eta_1(\tau_j)}$, we expect the first interaction to require a duration different from the following interactions, all of which may be chosen to be the same. Also, we see from our graphs that $\abs{\gamma_1(\tau_1)\,\gamma_2(\tau_1)}$ and $\abs{\gamma_1(\tau_j)\,\eta_1(\tau_j)}$ are periodic. For the numerical simulations that follow, we will take the durations that give the first maximum.

\section{\label{sec:numerical}Results}
\begin{figure}[t!]
\includegraphics[scale=0.38]{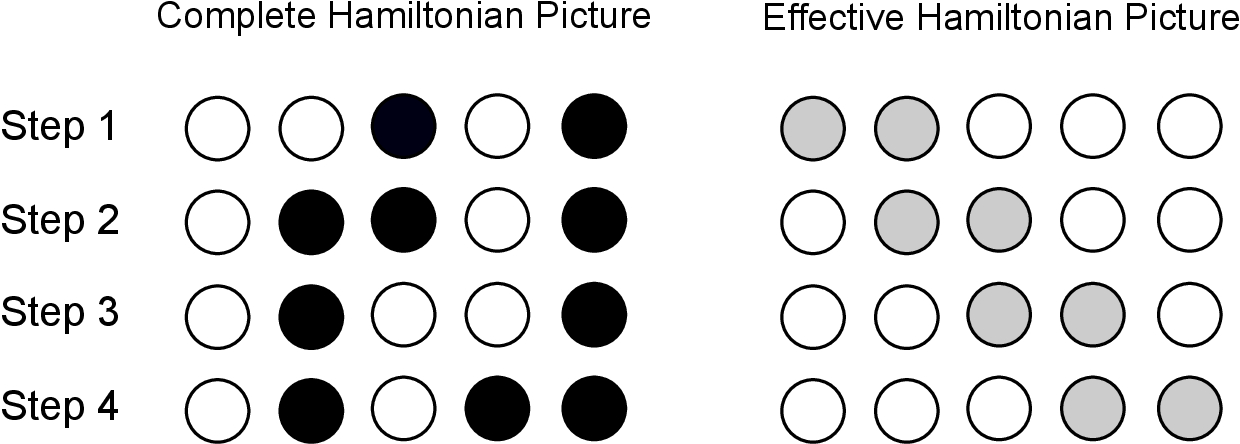}
\vspace*{6mm}
\caption{The diagram shows the general scheme we use to entangle the ends of a spin chain with $N=5$. In the complete Hamiltonian picture, a white circle at site $i$ represents a control field Hamiltonian at site $i$ described by the constants $n_x, n_y$ and a black circle at site $i$ represents a control field Hamiltonian at site $i$ described by the constants $m_x, m_y$. Notice that at each step, the two neighboring spins that are interacting have the same control field Hamiltonian at their sites, while the control field Hamiltonian for the rest of the chain is staggered. In the effective Hamiltonian picture, the two gray circles in each step represent the two-qubit interaction $\Bar{H}_i$ [see  Eq.~\eqref{eq:b}] that is caused by the same control field Hamiltonians at those sites. The white circles represent no interaction, which is caused by staggered fields at those sites.}
\label{fig:drawing}
\end{figure}

To test our predictions of perfect entanglement for the XY spin chain, we now perform numerical simulations. To measure the degree of entanglement, we calculate the concurrence between the two spins \cite{c40}. We calculate the partial trace of the density matrix of the system over all spins other than the two spins whose concurrence we aim to find and refer to it as $\rho_{2}$. We then calculate 
\begin{equation}
R = \sqrt{\sqrt{\rho_2} \widetilde{\rho}_2 \sqrt{\rho_2}},
\end{equation}
where $\widetilde{\rho}_2 = (\sigma_y \otimes \sigma_y)\rho_{2}^{\ast}(\sigma_y \otimes \sigma_y)$. The concurrence is given by
\begin{equation}
C(\rho) = \text{max}(0, \lambda_1 - \lambda_2 - \lambda_3 - \lambda_4),
\end{equation}
where $\lambda_1, \lambda_2, \lambda_3$, and $\lambda_4$ are the eigenvalues of the $R$ in descending order.
We evolve the spin chain in two different ways. First, we apply control fields to a spin chain that interacts with the environment and plot the concurrence between two spins as a function of time. We refer to this as the complete Hamiltonian picture, with the complete Hamiltonian given by $H_{0} + H_{\text{SB}} + H_{c}(t)$. Second, we show that the dynamics can be reproduced using the time-averaged effective Hamiltonian $\Bar{H}$. We refer to this as the effective Hamiltonian picture. In both approaches, we use the scheme discussed previously. That is, we make spins 1 and 2 interact until they are entangled and then turn off the interaction by switching the control fields. Then we make spins 2 and 3 interact until spins 1 and 3 are entangled and then turn off the interaction. We continue this process until we reach the spin that we wish to entangle with the first spin. Fig.~\ref{fig:drawing} illustrates the scheme for a chain with five spins.

\begin{figure}[t!]
  \includegraphics[width=.9\columnwidth]{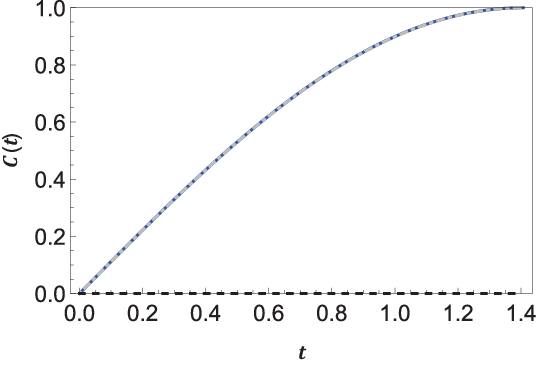}  
  \caption{Step 1. Spins 1 and 2 in an XY chain ($N=5, \zeta_{j,k}=1$) interact for $\tau_1$ ($\approx 1.4$) to become perfectly entangled, after which the interaction is effectively turned off. The concurrence between spins 1 and 2 is given by $C(t)$. The solid blue curve is the concurrence using the effective Hamiltonian, the dotted-dashed gray curve is the concurrence using the complete Hamiltonian, and the dashed black curve is the concurrence without the control fields. We use dimensionless units throughout with $\hbar = 1$.}
  \label{fig:N5step1}
\end{figure}
\begin{figure}[t!]
   \includegraphics[width=.9\columnwidth]{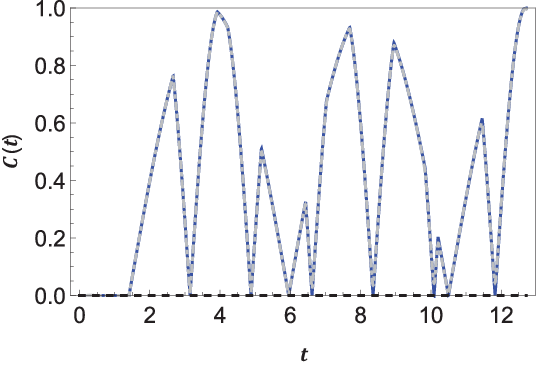}
   \caption{Step 2. Spins 2 and 3 interact for $\tau_2$ ($\approx 11.3$), after which spins 1 and 3 are perfectly entangled, as shown by their concurrence $C(t)$. Notice that $C(t) = 0$ for the duration $\tau_1$ since that is the duration for which spins $1$ and $2$ were interacting while all other spins were decoupled from each other.}
   \label{fig:N5step2}
\end{figure}
\begin{figure}[t!]
    \includegraphics[width=.9\columnwidth]{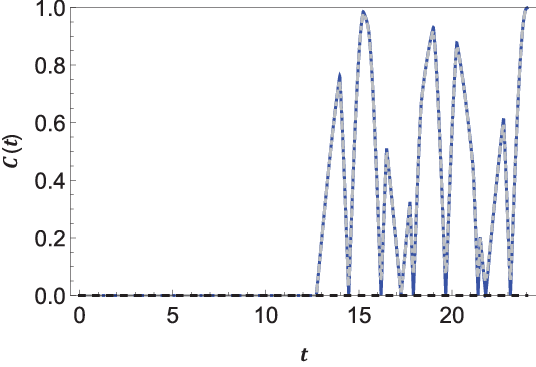} 
    \caption{Step 3. Spins 3 and 4 interact for $\tau_3$ ($\approx 11.3$), after which spins 1 and 4 are perfectly entangled, as shown by their concurrence $C(t)$. $C(t)$ begins to increase after $(\tau_1 + \tau_2)$ since that is the duration during which the first two steps of the scheme take place.}
   \label{fig:N5step3}
\end{figure}

\begin{figure}[t!]
   \centering
   \includegraphics[width=.9\columnwidth]{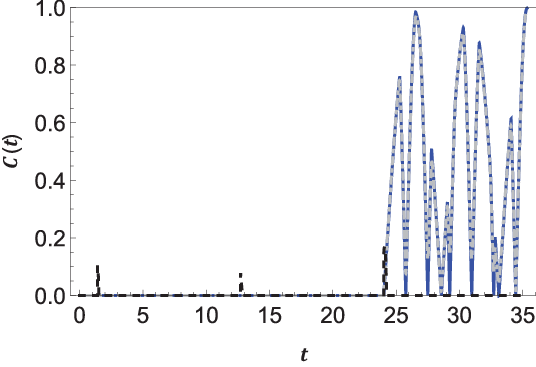}  
   \caption{Step 4. Spins 4 and 5 interact for $\tau_4$ ($\approx 11.3$), after which spins 1 and 5 are perfectly entangled, as shown by their concurrence $C(t)$. We have checked that when $C(t)$ is approximately one, the state of spins $1$ and $5$ is very close to $\frac{1}{\sqrt{2}}(\ket{00} + i\ket{11})$, which is a perfectly entangled state.}
  \label{fig:N5step4}
\end{figure}

\begin{figure}[t!]
   \includegraphics[width=.9\columnwidth]{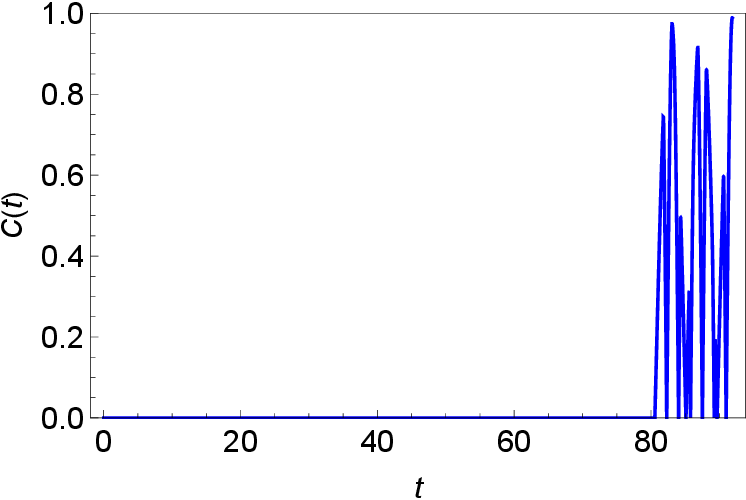}  
   \caption{$C(t)$ is the concurrence between spins $1$ and $10$ in an XY chain ($N=10, \zeta_{jk} = 1$). From $t=0$ to $t\approx 80$, consecutive pairs of neighboring spins interact. Thereafter, spins 9 and 10 are made to effectively interact until spins 1 and 10 are perfectly entangled.}
   \label{fig:N10}
\end{figure}

We now present our numerical results illustrating the entanglement at the end of each step. Recall that the spin chain is initialized in the state $\prod_{i=1}^{N} \ket{0}_{i}$. The noise terms $B_{k}^{(j)}$ in the expression for $H_{SB}$ [see Eq.~\eqref{eq:HSB}] are taken to be the same for each site for simplicity. These noise terms are generated via independent Ornstein-Uhlenbeck processes with zero mean, correlation time $\tau = 0.5$, and standard deviation $\sigma = 2.0$. Results for the concurrence are presented in Figs.~\ref{fig:N5step1}, \ref{fig:N5step2}, \ref{fig:N5step3}, and \ref{fig:N5step4}. Each plot shows the concurrence $C(t)$ between the first spin and the most recent spin to have experienced an interaction with its neighboring spin. The solid, blue curve shows the concurrence using the effective Hamiltonian, the dot-dashed gray curve shows the concurrence using the complete Hamiltonian, and the black, dashed curve shows the concurrence using only $H_{0} + H_{\text{SB}}$ (that is, no control fields are applied). Due to the noise, it is not surprising that the black, dashed curve largely overlaps with the horizontal axis. The overlap of the solid blue and dot-dashed grey curves shows that the effective Hamiltonian approach captures the exact dynamics very well. Notice that by the end of step 4, the concurrence between the first and last spins is practically one. We have also checked that the state of spins $1$ and $5$ at the end of step 4 is very close to $\frac{1}{\sqrt{2}} (\ket{00} + i\ket{11})$. For $N > 5$, simulating the chain via the complete Hamiltonian requires long durations. Since we have already shown the equivalence of the complete and effective Hamiltonian pictures, we simply use the effective Hamiltonian to simulate longer spin chains. Our scheme works, as expected, for a longer spin chain as well [see Fig.~\ref{fig:N10}].

\begin{figure}[t!]
   \includegraphics[width=.9\columnwidth]{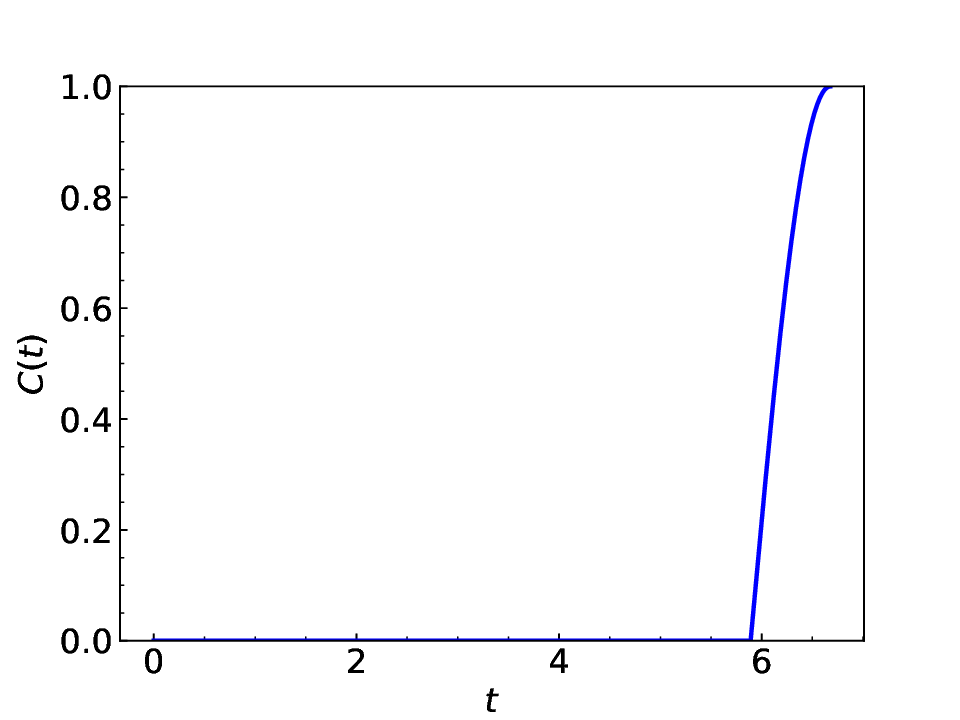}  
   \caption{$C(t)$ is the concurrence between spins $1$ and $10$ for the XXX model with $N = 10$. From $t=0$ to $t\approx 5.9$, consecutive pairs of neighboring spins interact. From $t\approx 5.9$ onwards, spins 9 and 10 are made to interact until spins 1 and 10 are perfectly entangled.}
   \label{fig:xyziso}
\end{figure}

The entanglement scheme we have presented is not just restricted to the XY model. A necessary condition for our scheme to work is that the effective Hamiltonian must be able to generate perfect entanglement between two neighboring spins, given a suitable initial state. Let us then examine the isotropic XYZ model (or the XXX model) with $\zeta_{j1} = \zeta_{j2} = \zeta_{j3} = 1$ in Eq.~\eqref{eq:H0}. We now choose the initial state of the spin chain to be 
\begin{equation*}
    \ket{\psi(0)} =  \ket{10}_{1,2}\otimes \ket{\bm{0}}.
\end{equation*}
The effective Hamiltonian describing the spin-spin interactions between two spins when the same control fields are applied to the two spins is now 
\begin{equation}
\Bar{H}_i = \sigma_{x}^{(i)}\sigma_{x}^{(i+1)} + \sigma_{y}^{(i)}\sigma_{y}^{(i+1)}  + \sigma_{z}^{(i)}\sigma_{z}^{(i+1)} . 
\end{equation}
As in Sec.~\ref{sec:scheme}, we find the evolution due to this effective interaction by finding the eigenstates of the effective Hamiltonian in the $\{\ket{00}_{i,i+1}$, $\ket{01}_{i,i+1}$, $\ket{10}_{i,i+1}$, $\ket{11}_{i,i+1}\}$ basis. These very familiar eigenstates are
 \begin{gather*}
    \ket{e_1} = \frac{1}{\sqrt{2}}
    \begin{bmatrix}
    0  \\ 1 \\ -1\\ 0
    \end{bmatrix},\,\,
    \ket{e_2} = 
    \begin{bmatrix}
    1\\ 0 \\ 0 \\ 0
    \end{bmatrix},\,\,\\
    \ket{e_3} = \frac{1}{\sqrt{2}}
    \begin{bmatrix}
    0 \\ 1 \\1 \\ 0
    \end{bmatrix},\,\,
    \ket{e_4} =
    \begin{bmatrix}
    0 \\ 0 \\ 0 \\ 1
    \end{bmatrix},
\end{gather*}
with eigenvalues $\lambda_1 = -3, \lambda_2 = 0, \lambda_3=1 \text{ and } \lambda_4=1$ respectively. We can then also write
\begin{align*}
\ket{00}_{i,i+1} &= \ket{e_2}, \\
\ket{01}_{i,i+1} &= -\frac{1}{\sqrt{2}} \ket{e_1} + \frac{1}{\sqrt{2}} \ket{e_3},\\ 
\ket{10}_{i,i+1} &= \frac{1}{\sqrt{2}} \ket{e_1} + \frac{1}{\sqrt{2}} \ket{e_3}, \\
\ket{11}_{i,i+1} &= \ket{e_4}.
\end{align*}
We now let the first two spins interact. After time $t$, 
\begin{equation}
\ket{\psi(t)} = \left(\chi_{1}(t) \ket{10}_{1,2} + \chi_{2}(t) \ket{01}_{1,2}\right)\ket{\mathbf{0}},
\end{equation}
where 
\begin{align}
\chi_{1}(t) &= \frac{1}{2}\Bigl( e^{-it} + e^{i3t} \Bigr), \\
\chi_{2}(t) &= \frac{1}{2}\Bigl( e^{-it} - e^{i3t} \Bigr).
\end{align}
As before, we select a time $\tau_1$ such that spins $1$ and $2$ are perfectly entangled. Then, we let the next spins interact pairwise for time $\tau_j$. We arrive at a similar condition to Eq.~\eqref{eq:det} for spins $1$ and $i + 1$ to be fully entangled:
\begin{equation}
\abs{\chi_1(\tau_{1})\,\chi_2(\tau_{1})\prod_{j=2}^{i}\chi_2(\tau_{j})} = 0.5.
\label{eq:isotropicdet}
\end{equation}
Note that this is a somewhat simpler condition as compared to Eq.~\eqref{eq:det} since we do not have to consider the evolution of $\ket{00}_{j,j+1}$ since it is an eigenstate of the effective Hamiltonian. To fulfill this condition, we can choose $\abs{\chi_1(\tau_1)\chi_2(\tau_1)} = 0.5$ and $\abs{\chi_2(\tau_j)} = 1$ with all $\tau_j$ equal to $\tau_2$. We can work out the required values of $\tau_1$ and $\tau_2$ by first finding that $\abs{\chi_1(\tau_1)\chi_2(\tau_1)} = \frac{1}{2} \sin(4\tau_1)$ and $\abs{\chi_2(\tau_2)} = \sin(2\tau_2)$. It then follows that we can choose $\tau_1 = \frac{\pi}{8}$ and $\tau_2 = \frac{\pi}{4}$. With these times, spins 1 and $i + 1$ are in a Bell state, while all the other spins are in the state $\ket{0}$. Fig.~\ref{fig:xyziso} is an illustration of the first and last spins of a XX spin chain being entangled via our scheme. 

\begin{figure}[h!]
    \centering
    \includegraphics[width=.9\columnwidth]{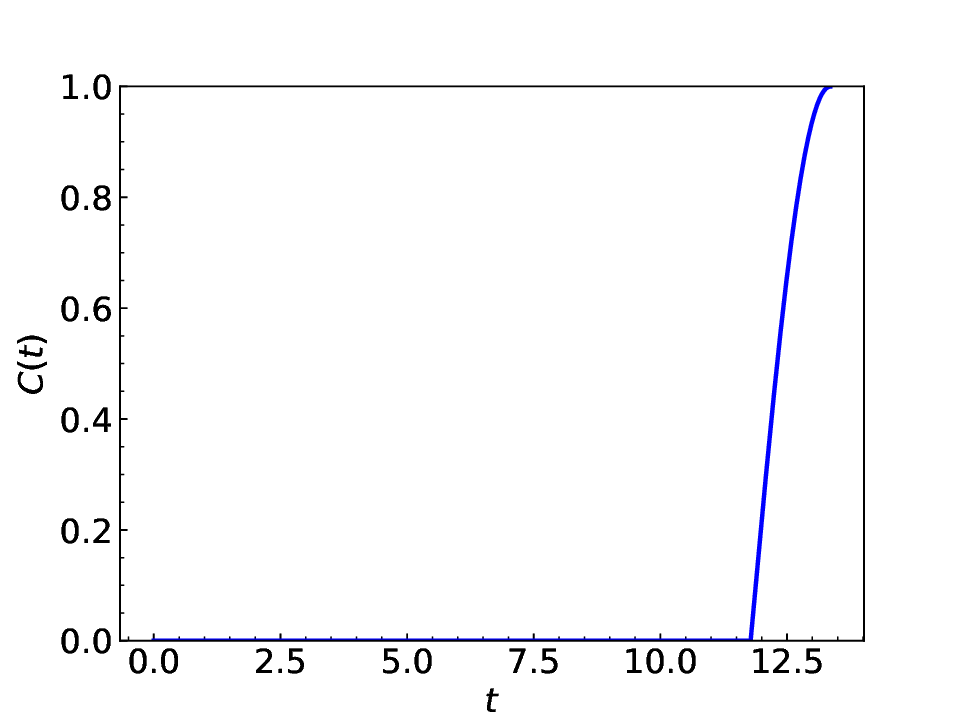}
    \caption{$C(t)$ is the concurrence between spins $1$ and $10$ for the quantum Ising model with $N = 10$ following our scheme.}
    \label{figising}
\end{figure}

Let us now consider the quantum Ising spin chain with $\zeta_{j1} = 1$ and $\zeta_{j2} = \zeta_{j3} = 0$. The effective Hamiltonian in this case is
\begin{equation}
    \Bar{H}_i = \frac{1}{2}\left
[ \sigma_{x}^{(i)}\sigma_{x}^{(i+1)} + \sigma_{z}^{(i)}\sigma_{z}^{(i+1)}\right].
    \end{equation}
This is really the usual XX interaction since, by performing a rotation of the coordinate axes, we can write the effective Hamiltonian in terms of $\sigma_y$ instead. That is,
\begin{equation}
    \Bar{H}'_i = \frac{1}{2} \left
(\sigma_{x}^{i}\sigma_{x}^{i+1} + \sigma_{y}^{i}\sigma_{y}^{i+1} \right).
    \end{equation}
In this rotated basis, we choose the initial state to be $\ket{10}_{1,2}\otimes\ket{\mathbf{0}}$. The same eigenbasis $\ket{e_i}$ is the same as that for the XXX model with eigenvalues $\lambda_1 = -1, \lambda_2 =0, \lambda_3 = 1, \lambda_4 =0$. Letting the first two spins interact we have
\begin{equation}
\ket{\psi(t)} = \left(\alpha_{1}(t) \ket{10}_{1,2} + \alpha_{2}(t) \ket{01}_{1,2}\right)\ket{\mathbf{0}},
\end{equation}
where 
\begin{align}
\alpha_{1}(t) &= \frac{1}{2}\Bigl( e^{-it} + e^{it} \Bigr) = \cos(t),  \\
\alpha_{2}(t) &= \frac{1}{2}\Bigl( e^{-it} - e^{it} \Bigr) = -i\sin(t).
\end{align}
To get spins 1 and $i + 1$ in a maximally entangled state, we now have the condition
\begin{equation}
\abs{\alpha_1(\tau_{1})\,\alpha_2(\tau_{1})\prod_{j=2}^{i}\alpha_2(\tau_{j})} = 0.5,
\end{equation}
To fulfill this, we can choose $\tau_1$ such that $\abs{\cos(\tau_1)\sin(\tau_1)} = 0.5$, and setting all the $\tau_j$ to be equal to $\tau_2$ such that $\abs{\sin(\tau_2)} = 1$. With these times, spins 1 and $i + 1$ are in a Bell state and all the others spins are in the state $\ket{0}$. The generation of the entanglement is illustrated in Fig.~\ref{figising}.

\begin{figure}[h!]
    \centering
    \includegraphics[width=.9\columnwidth]{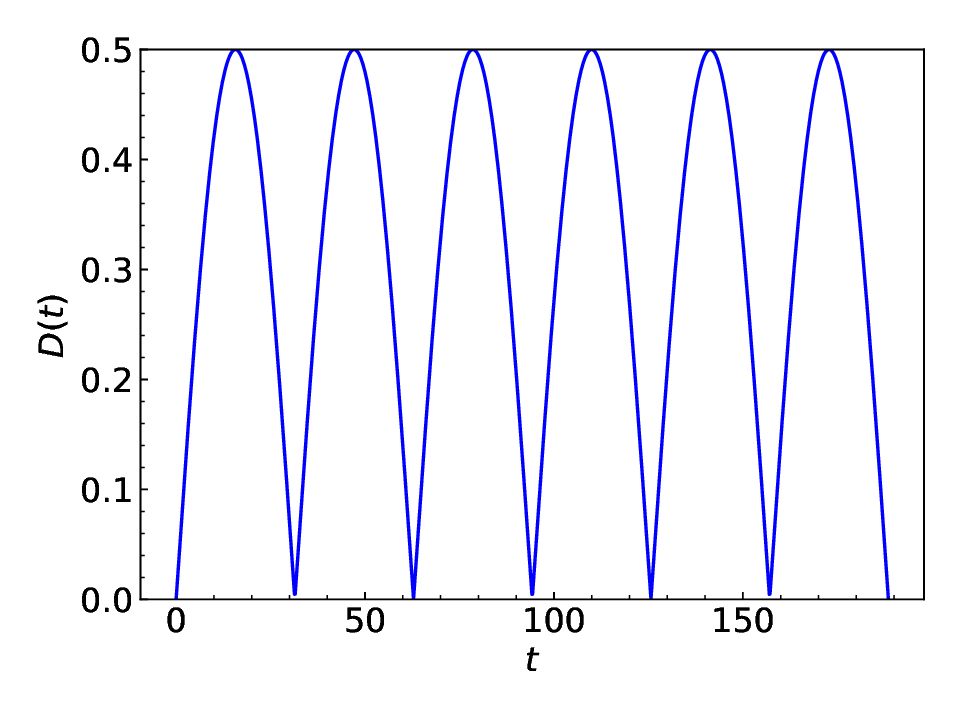}
    \caption{A plot of $D(t)$ against time, where $D(t) = \abs{\mu_1(t)\,\mu_2(t)}$.}
    \label{fig:det3}
\end{figure}

\begin{figure}[h!]
    \centering
    \includegraphics[width=.9\columnwidth]{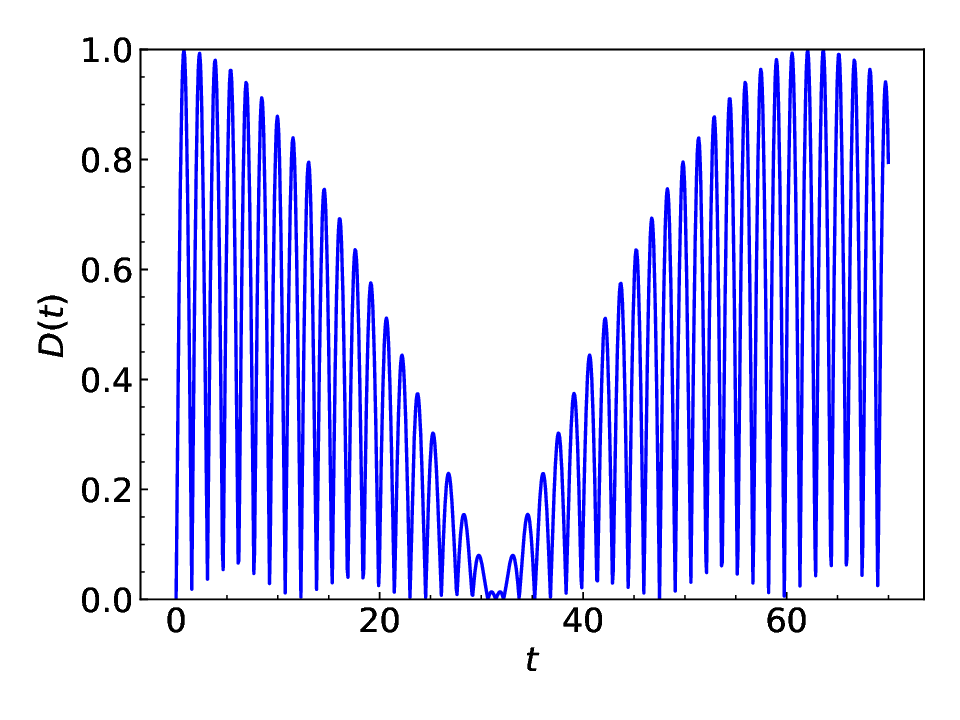}
    \caption{A plot of $D(t)$ against time, where $D(t) = \abs{\mu_1(t)\,\nu_1(t)}$.}
    \label{fig:det4}
\end{figure}

\begin{figure}[h!]
   \includegraphics[width=.9\columnwidth]{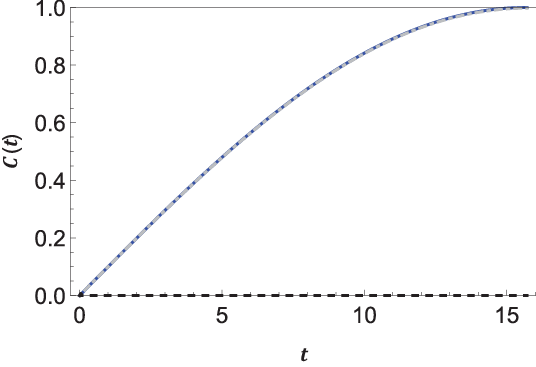} 
   \centering 
   \caption{Step 1. Spins 1 and 2 in an anisotropic XYZ chain ($N=5, \zeta_{j1} = 1.1, \zeta_{j2} = 1, \zeta_{j3} = 1$) interact for $\tau_1$ ($\approx 15.7$).}
   \label{fig:N5step1XYZ}
\end{figure}

\begin{figure}[htb!]
   \includegraphics[width=.9\columnwidth]{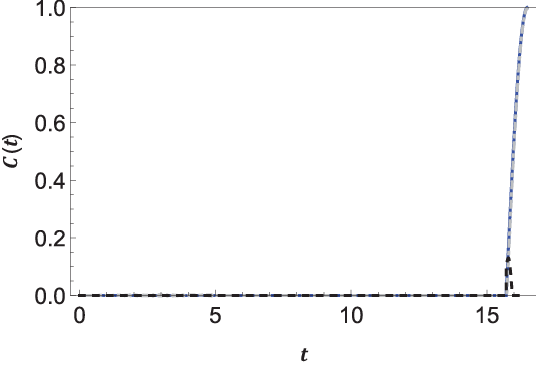}  
   \centering 
   \caption{Step 2. Spins 2 and 3 interact for $\tau_2$ ($\approx 0.77$). $C(t)$ is the concurrence between spins $1$ and $3$.}
   \label{fig:N5step2XYZ}
\end{figure}

\begin{figure}[htb!]
   \includegraphics[width=.9\columnwidth]{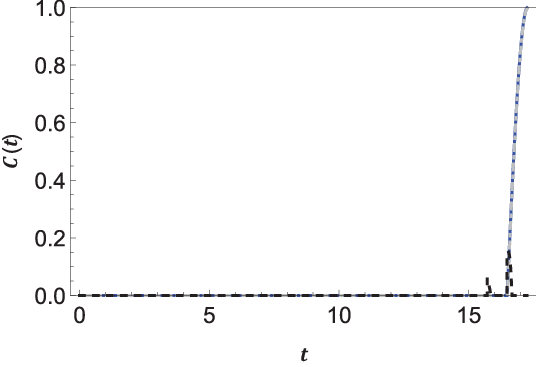} 
   \centering
   \caption{Step 3. Spins 3 and 4 interact for $\tau_3$ ($\approx 0.77$). $C(t)$ is the concurrence between spins $1$ and $4$.}
   \label{fig:N5step3XYZ}
\end{figure}

\begin{figure}[htb!]
   \includegraphics[width=.9\columnwidth]{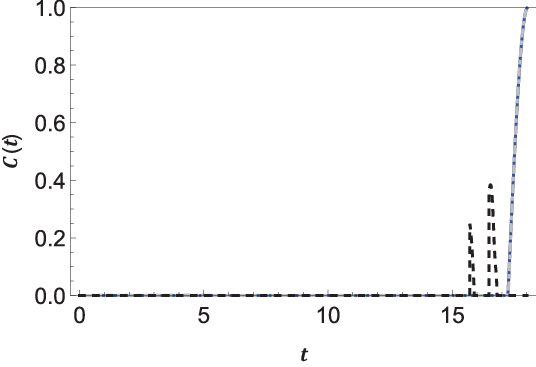}  
   \centering
   \caption{Step 4. Spins 4 and 5 interact for $\tau_4$ ($\approx 0.77$). $C(t)$ is the concurrence between spins $1$ and $5$. At the point when the interaction stops, $C(t)=0.997$ rounded off to three decimal places.}
   \label{fig:N5step4XYZ}
\end{figure}

Finally, let us consider an anisotropic XYZ model. We take $\zeta_{j1} = 1.1, \zeta_{j2} = 1, \zeta_{j3} = 1$. By now, it should be obvious how to proceed with our scheme. The effective Hamiltonian is
\begin{equation}
\Bar{H}_i = 1.05[\sigma_{x}^{(i)}\sigma_{x}^{(i+1)}+ \sigma_{z}^{(i)}\sigma_{z}^{(i+1)}] + \sigma_{y}^{(i)}\sigma_{y}^{(i+1)}.
\end{equation}
We choose the initial state to be $\ket{\mathbf{0}}$. Spins 1 and $i + 1$ are now maximally entangled if 
\begin{equation}\label{eq:condition}
\abs{\mu_1(\tau_{1})\,\mu_2(\tau_{1})\Bigl(\prod_{j=2}^{i}\mu_1(\tau_{j})\,\nu_1(\tau_{j})\Bigr)} = 0.5,
\end{equation}
where 
\begin{align}
\mu_1 &= \frac{1}{2}\Bigl( e^{-i \lambda_4 t} + e^{-i \lambda_2 t} \Bigr), \\ 
\mu_2 &= \frac{1}{2}\Bigl( e^{-i \lambda_4 t} - e^{-i \lambda_2 t} \Bigr),  \\
\nu_1 &= \frac{1}{2}\Bigl( e^{-i \lambda_1 t} - e^{-i \lambda_3 t} \Bigr),
\end{align}
and $\lambda_1 = -3.1,~ \lambda_2 = 1, ~\lambda_3=1 \text{ and } \lambda_4= 1.1$. To fulfill this condition, we can set $\abs{\mu_1(\tau_{1})\,\mu_2(\tau_{1})} = 0.5$, and $\abs{\mu_1(\tau_{2})\,\nu_1(\tau_{2})} = 1$, where all the $\tau_j$ have been set equal to $\tau_2$. Figs.~\ref{fig:det3} and \ref{fig:det4} show that these two conditions can indeed be satisfied.  
Proceeding further, Figs.~\ref{fig:N5step1XYZ}, \ref{fig:N5step2XYZ}, \ref{fig:N5step3XYZ}, and \ref{fig:N5step4XYZ} show how entanglement is generated between the first spin and the rest of the spins one by one. As before, the dot-dashed gray curve shows the concurrence obtained by using the complete Hamiltonian, the solid, blue curve are the results with the effective Hamiltonian approach, and the black, dashed curve is the concurrence in the absence of any control fields. These plots not only illustrate the validity of the effective Hamiltonian approach with highly efficient removal of the effect of the environment, but also the fact that maximal bipartite entanglement can be generated between two spins in the spin chain. 


\section{\label{sec:conclusion}Conclusion}
In conclusion, we have shown that by applying staggered fields to a spin chain, not only can we largely decouple the spin chain from the environment, but we can also suppress the spin-spin interactions, effectively obtaining a chain of non-interacting spins. Then, by considering a combination of constant and staggered configurations of strong static and oscillating fields, we demonstrated how interactions between two spins in the chain can be selectively induced. By diagonalizing the effective two-qubit interaction Hamiltonian, we evolved the XX, XXX, Ising, and the anisotropic XYZ spin chains under a series of interactions that allowed us to generate maximal entanglement between any two spins in the spin chain. Our results are interesting given the significance of entanglement in spin chains, and the importance of protecting the entanglement from the environment when considering applications in quantum computation and information. Our proposed scheme can potentially lead to the generation of near-perfect entanglement between the far ends of a long spin chain, even in the presence of significant external noise.


\begin{thebibliography}{38}%
\makeatletter
\providecommand \@ifxundefined [1]{%
 \@ifx{#1\undefined}
}%
\providecommand \@ifnum [1]{%
 \ifnum #1\expandafter \@firstoftwo
 \else \expandafter \@secondoftwo
 \fi
}%
\providecommand \@ifx [1]{%
 \ifx #1\expandafter \@firstoftwo
 \else \expandafter \@secondoftwo
 \fi
}%
\providecommand \natexlab [1]{#1}%
\providecommand \enquote  [1]{``#1''}%
\providecommand \bibnamefont  [1]{#1}%
\providecommand \bibfnamefont [1]{#1}%
\providecommand \citenamefont [1]{#1}%
\providecommand \href@noop [0]{\@secondoftwo}%
\providecommand \href [0]{\begingroup \@sanitize@url \@href}%
\providecommand \@href[1]{\@@startlink{#1}\@@href}%
\providecommand \@@href[1]{\endgroup#1\@@endlink}%
\providecommand \@sanitize@url [0]{\catcode `\\12\catcode `\$12\catcode
  `\&12\catcode `\#12\catcode `\^12\catcode `\_12\catcode `\%12\relax}%
\providecommand \@@startlink[1]{}%
\providecommand \@@endlink[0]{}%
\providecommand \url  [0]{\begingroup\@sanitize@url \@url }%
\providecommand \@url [1]{\endgroup\@href {#1}{\urlprefix }}%
\providecommand \urlprefix  [0]{URL }%
\providecommand \Eprint [0]{\href }%
\providecommand \doibase [0]{https://doi.org/}%
\providecommand \selectlanguage [0]{\@gobble}%
\providecommand \bibinfo  [0]{\@secondoftwo}%
\providecommand \bibfield  [0]{\@secondoftwo}%
\providecommand \translation [1]{[#1]}%
\providecommand \BibitemOpen [0]{}%
\providecommand \bibitemStop [0]{}%
\providecommand \bibitemNoStop [0]{.\EOS\space}%
\providecommand \EOS [0]{\spacefactor3000\relax}%
\providecommand \BibitemShut  [1]{\csname bibitem#1\endcsname}%
\let\auto@bib@innerbib\@empty
\bibitem [{\citenamefont {Kashurnikov}\ \emph {et~al.}(1999)\citenamefont
  {Kashurnikov}, \citenamefont {Prokof'ev}, \citenamefont {Svistunov},\ and\
  \citenamefont {Troyer}}]{c1}%
  \BibitemOpen
  \bibfield  {author} {\bibinfo {author} {\bibfnamefont {V.~A.}\ \bibnamefont
  {Kashurnikov}}, \bibinfo {author} {\bibfnamefont {N.~V.}\ \bibnamefont
  {Prokof'ev}}, \bibinfo {author} {\bibfnamefont {B.~V.}\ \bibnamefont
  {Svistunov}},\ and\ \bibinfo {author} {\bibfnamefont {M.}~\bibnamefont
  {Troyer}},\ }\bibfield  {title} {\bibinfo {title} {Quantum spin chains in a
  magnetic field},\ }\href {https://doi.org/10.1103/PhysRevB.59.1162}
  {\bibfield  {journal} {\bibinfo  {journal} {Phys. Rev. B}\ }\textbf {\bibinfo
  {volume} {59}},\ \bibinfo {pages} {1162} (\bibinfo {year}
  {1999})}\BibitemShut {NoStop}%
\bibitem [{\citenamefont {Marchukov}\ \emph {et~al.}(2016)\citenamefont
  {Marchukov}, \citenamefont {Volosniev}, \citenamefont {Valiente},
  \citenamefont {Petrosyan},\ and\ \citenamefont {Zinner}}]{c2}%
  \BibitemOpen
  \bibfield  {author} {\bibinfo {author} {\bibfnamefont {O.~V.}\ \bibnamefont
  {Marchukov}}, \bibinfo {author} {\bibfnamefont {A.~G.}\ \bibnamefont
  {Volosniev}}, \bibinfo {author} {\bibfnamefont {M.}~\bibnamefont {Valiente}},
  \bibinfo {author} {\bibfnamefont {D.}~\bibnamefont {Petrosyan}},\ and\
  \bibinfo {author} {\bibfnamefont {N.~T.}\ \bibnamefont {Zinner}},\ }\bibfield
   {title} {\bibinfo {title} {Quantum spin transistor with a heisenberg spin
  chain},\ }\href {https://doi.org/10.1038/ncomms13070} {\bibfield  {journal}
  {\bibinfo  {journal} {Nat. Commun.}\ }\textbf {\bibinfo {volume} {7}},\
  \bibinfo {pages} {13070} (\bibinfo {year} {2016})}\BibitemShut {NoStop}%
\bibitem [{\citenamefont {Gubin}\ and\ \citenamefont {F.~Santos}(2012)}]{c3}%
  \BibitemOpen
  \bibfield  {author} {\bibinfo {author} {\bibfnamefont {A.}~\bibnamefont
  {Gubin}}\ and\ \bibinfo {author} {\bibfnamefont {L.}~\bibnamefont
  {F.~Santos}},\ }\bibfield  {title} {\bibinfo {title} {Quantum chaos: An
  introduction via chains of interacting spins 1/2},\ }\href
  {https://doi.org/10.1119/1.3671068} {\bibfield  {journal} {\bibinfo
  {journal} {Am. J. Phys.}\ }\textbf {\bibinfo {volume} {80}},\ \bibinfo
  {pages} {246} (\bibinfo {year} {2012})} \BibitemShut {NoStop}%
\bibitem [{\citenamefont {Anderson}(1958)}]{c4}%
  \BibitemOpen
  \bibfield  {author} {\bibinfo {author} {\bibfnamefont {P.~W.}\ \bibnamefont
  {Anderson}},\ }\bibfield  {title} {\bibinfo {title} {Absence of diffusion in
  certain random lattices},\ }\href {https://doi.org/10.1103/PhysRev.109.1492}
  {\bibfield  {journal} {\bibinfo  {journal} {Phys. Rev.}\ }\textbf {\bibinfo
  {volume} {109}},\ \bibinfo {pages} {1492} (\bibinfo {year}
  {1958})}\BibitemShut {NoStop}%
\bibitem [{\citenamefont {Spiller}\ \emph {et~al.}(2007)\citenamefont
  {Spiller}, \citenamefont {D'Amico},\ and\ \citenamefont
  {Lovett}}]{SpillerNJP2007}%
  \BibitemOpen
  \bibfield  {author} {\bibinfo {author} {\bibfnamefont {T.~P.}\ \bibnamefont
  {Spiller}}, \bibinfo {author} {\bibfnamefont {I.}~\bibnamefont {D'Amico}},\
  and\ \bibinfo {author} {\bibfnamefont {B.~W.}\ \bibnamefont {Lovett}},\
  }\bibfield  {title} {\bibinfo {title} {Entanglement distribution for a
  practical quantum-dot-based quantum processor architecture},\ }\href
  {https://doi.org/10.1088/1367-2630/9/1/020} {\bibfield  {journal} {\bibinfo
  {journal} {New Journal of Physics}\ }\textbf {\bibinfo {volume} {9}},\
  \bibinfo {pages} {20} (\bibinfo {year} {2007})}\BibitemShut {NoStop}%
\bibitem [{\citenamefont {Banchi}\ \emph {et~al.}(2011)\citenamefont {Banchi},
  \citenamefont {Bayat}, \citenamefont {Verrucchi},\ and\ \citenamefont
  {Bose}}]{LeonardoPRL2011}%
  \BibitemOpen
  \bibfield  {author} {\bibinfo {author} {\bibfnamefont {L.}~\bibnamefont
  {Banchi}}, \bibinfo {author} {\bibfnamefont {A.}~\bibnamefont {Bayat}},
  \bibinfo {author} {\bibfnamefont {P.}~\bibnamefont {Verrucchi}},\ and\
  \bibinfo {author} {\bibfnamefont {S.}~\bibnamefont {Bose}},\ }\bibfield
  {title} {\bibinfo {title} {Nonperturbative entangling gates between distant
  qubits using uniform cold atom chains},\ }\href
  {https://doi.org/10.1103/PhysRevLett.106.140501} {\bibfield  {journal}
  {\bibinfo  {journal} {Phys. Rev. Lett.}\ }\textbf {\bibinfo {volume} {106}},\
  \bibinfo {pages} {140501} (\bibinfo {year} {2011})}\BibitemShut {NoStop}%
\bibitem [{\citenamefont {Rafiee}\ and\ \citenamefont
  {Mokhtari}(2013)}]{HosseinPRA2013}%
  \BibitemOpen
  \bibfield  {author} {\bibinfo {author} {\bibfnamefont {M.}~\bibnamefont
  {Rafiee}}\ and\ \bibinfo {author} {\bibfnamefont {H.}~\bibnamefont
  {Mokhtari}},\ }\bibfield  {title} {\bibinfo {title} {Long-distance
  entanglement generation by local rotational protocols in spin chains},\
  }\href {https://doi.org/10.1103/PhysRevA.87.022304} {\bibfield  {journal}
  {\bibinfo  {journal} {Phys. Rev. A}\ }\textbf {\bibinfo {volume} {87}},\
  \bibinfo {pages} {022304} (\bibinfo {year} {2013})}\BibitemShut {NoStop}%
\bibitem [{\citenamefont {Sahling}\ \emph {et~al.}(2015)\citenamefont
  {Sahling}, \citenamefont {Remenyi}, \citenamefont {Paulsen}, \citenamefont
  {Monceau}, \citenamefont {Saligrama}, \citenamefont {Marin}, \citenamefont
  {Revcolevschi}, \citenamefont {Regnault}, \citenamefont {Raymond},\ and\
  \citenamefont {Lorenzo}}]{SahlingNatPhys2015}%
  \BibitemOpen
  \bibfield  {author} {\bibinfo {author} {\bibfnamefont {S.}~\bibnamefont
  {Sahling}}, \bibinfo {author} {\bibfnamefont {G.}~\bibnamefont {Remenyi}},
  \bibinfo {author} {\bibfnamefont {C.}~\bibnamefont {Paulsen}}, \bibinfo
  {author} {\bibfnamefont {P.}~\bibnamefont {Monceau}}, \bibinfo {author}
  {\bibfnamefont {V.}~\bibnamefont {Saligrama}}, \bibinfo {author}
  {\bibfnamefont {C.}~\bibnamefont {Marin}}, \bibinfo {author} {\bibfnamefont
  {A.}~\bibnamefont {Revcolevschi}}, \bibinfo {author} {\bibfnamefont {L.~P.}\
  \bibnamefont {Regnault}}, \bibinfo {author} {\bibfnamefont {S.}~\bibnamefont
  {Raymond}},\ and\ \bibinfo {author} {\bibfnamefont {J.~E.}\ \bibnamefont
  {Lorenzo}},\ }\bibfield  {title} {\bibinfo {title} {Experimental realization
  of long-distance entanglement between spins in antiferromagnetic quantum spin
  chains},\ }\href {https://doi.org/10.1038/nphys3186} {\bibfield  {journal}
  {\bibinfo  {journal} {Nature Phys.}\ }\textbf {\bibinfo {volume} {11}},\
  \bibinfo {pages} {255} (\bibinfo {year} {2015})}\BibitemShut {NoStop}%
\bibitem [{\citenamefont {Estarellas}\ \emph {et~al.}(2017)\citenamefont
  {Estarellas}, \citenamefont {D'Amico},\ and\ \citenamefont
  {Spiller}}]{EstarellasPRA2017}%
  \BibitemOpen
  \bibfield  {author} {\bibinfo {author} {\bibfnamefont {M.~P.}\ \bibnamefont
  {Estarellas}}, \bibinfo {author} {\bibfnamefont {I.}~\bibnamefont
  {D'Amico}},\ and\ \bibinfo {author} {\bibfnamefont {T.~P.}\ \bibnamefont
  {Spiller}},\ }\bibfield  {title} {\bibinfo {title} {Robust quantum
  entanglement generation and generation-plus-storage protocols with spin
  chains},\ }\href {https://doi.org/10.1103/PhysRevA.95.042335} {\bibfield
  {journal} {\bibinfo  {journal} {Phys. Rev. A}\ }\textbf {\bibinfo {volume}
  {95}},\ \bibinfo {pages} {042335} (\bibinfo {year} {2017})}\BibitemShut
  {NoStop}%
\bibitem [{\citenamefont {D\"ur}\ \emph {et~al.}(2005)\citenamefont {D\"ur},
  \citenamefont {Hartmann}, \citenamefont {Hein}, \citenamefont {Lewenstein},\
  and\ \citenamefont {Briegel}}]{c10}%
  \BibitemOpen
  \bibfield  {author} {\bibinfo {author} {\bibfnamefont {W.}~\bibnamefont
  {D\"ur}}, \bibinfo {author} {\bibfnamefont {L.}~\bibnamefont {Hartmann}},
  \bibinfo {author} {\bibfnamefont {M.}~\bibnamefont {Hein}}, \bibinfo {author}
  {\bibfnamefont {M.}~\bibnamefont {Lewenstein}},\ and\ \bibinfo {author}
  {\bibfnamefont {H.-J.}\ \bibnamefont {Briegel}},\ }\bibfield  {title}
  {\bibinfo {title} {Entanglement in spin chains and lattices with long-range
  ising-type interactions},\ }\href
  {https://doi.org/10.1103/PhysRevLett.94.097203} {\bibfield  {journal}
  {\bibinfo  {journal} {Phys. Rev. Lett.}\ }\textbf {\bibinfo {volume} {94}},\
  \bibinfo {pages} {097203} (\bibinfo {year} {2005})}\BibitemShut {NoStop}%
\bibitem [{\citenamefont {Plastina}\ and\ \citenamefont
  {Apollaro}(2007)}]{c11}%
  \BibitemOpen
  \bibfield  {author} {\bibinfo {author} {\bibfnamefont {F.}~\bibnamefont
  {Plastina}}\ and\ \bibinfo {author} {\bibfnamefont {T.~J.~G.}\ \bibnamefont
  {Apollaro}},\ }\bibfield  {title} {\bibinfo {title} {Local control of
  entanglement in a spin chain},\ }\href
  {https://doi.org/10.1103/PhysRevLett.99.177210} {\bibfield  {journal}
  {\bibinfo  {journal} {Phys. Rev. Lett.}\ }\textbf {\bibinfo {volume} {99}},\
  \bibinfo {pages} {177210} (\bibinfo {year} {2007})}\BibitemShut {NoStop}%
\bibitem [{\citenamefont {Sheng}\ \emph {et~al.}(2008)\citenamefont {Sheng},
  \citenamefont {Deng},\ and\ \citenamefont {Zhou}}]{c12}%
  \BibitemOpen
  \bibfield  {author} {\bibinfo {author} {\bibfnamefont {Y.-B.}\ \bibnamefont
  {Sheng}}, \bibinfo {author} {\bibfnamefont {F.-G.}\ \bibnamefont {Deng}},\
  and\ \bibinfo {author} {\bibfnamefont {H.-Y.}\ \bibnamefont {Zhou}},\
  }\bibfield  {title} {\bibinfo {title} {Nonlocal entanglement concentration
  scheme for partially entangled multipartite systems with nonlinear optics},\
  }\href {https://doi.org/10.1103/PhysRevA.77.062325} {\bibfield  {journal}
  {\bibinfo  {journal} {Phys. Rev. A}\ }\textbf {\bibinfo {volume} {77}},\
  \bibinfo {pages} {062325} (\bibinfo {year} {2008})}\BibitemShut {NoStop}%
\bibitem [{\citenamefont {Austin}\ \emph {et~al.}(2019)\citenamefont {Austin},
  \citenamefont {Khan}, \citenamefont {Mudassar},\ and\ \citenamefont
  {Chaudhry}}]{c6}%
  \BibitemOpen
  \bibfield  {author} {\bibinfo {author} {\bibfnamefont {S.}~\bibnamefont
  {Austin}}, \bibinfo {author} {\bibfnamefont {M.~Q.}\ \bibnamefont {Khan}},
  \bibinfo {author} {\bibfnamefont {M.}~\bibnamefont {Mudassar}},\ and\
  \bibinfo {author} {\bibfnamefont {A.~Z.}\ \bibnamefont {Chaudhry}},\
  }\bibfield  {title} {\bibinfo {title} {Continuous dynamical decoupling of
  spin chains: Modulating the spin-environment and spin-spin interactions},\
  }\href {https://doi.org/10.1103/PhysRevA.100.022102} {\bibfield  {journal}
  {\bibinfo  {journal} {Phys. Rev. A}\ }\textbf {\bibinfo {volume} {100}},\
  \bibinfo {pages} {022102} (\bibinfo {year} {2019})}\BibitemShut {NoStop}%
\bibitem [{\citenamefont {Wanisch}\ and\ \citenamefont
  {Fritzsche}(2020)}]{c38}%
  \BibitemOpen
  \bibfield  {author} {\bibinfo {author} {\bibfnamefont {D.}~\bibnamefont
  {Wanisch}}\ and\ \bibinfo {author} {\bibfnamefont {S.}~\bibnamefont
  {Fritzsche}},\ }\bibfield  {title} {\bibinfo {title} {Driven spin chains as
  high-quality quantum routers},\ }\href
  {https://doi.org/10.1103/PhysRevA.102.032624} {\bibfield  {journal} {\bibinfo
   {journal} {Phys. Rev. A}\ }\textbf {\bibinfo {volume} {102}},\ \bibinfo
  {pages} {032624} (\bibinfo {year} {2020})}\BibitemShut {NoStop}%
\bibitem [{\citenamefont {Bazhanov}\ \emph {et~al.}(2018)\citenamefont
  {Bazhanov}, \citenamefont {Sivkov},\ and\ \citenamefont {Stepanyuk}}]{c7}%
  \BibitemOpen
  \bibfield  {author} {\bibinfo {author} {\bibfnamefont {D.~I.}\ \bibnamefont
  {Bazhanov}}, \bibinfo {author} {\bibfnamefont {I.~N.}\ \bibnamefont
  {Sivkov}},\ and\ \bibinfo {author} {\bibfnamefont {V.~S.}\ \bibnamefont
  {Stepanyuk}},\ }\bibfield  {title} {\bibinfo {title} {Engineering of
  entanglement and spin state transfer via quantum chains of atomic spins at
  large separations},\ }\href {https://doi.org/10.1038/s41598-018-32145-3}
  {\bibfield  {journal} {\bibinfo  {journal} {Sci. Rep.}\ }\textbf {\bibinfo
  {volume} {8}},\ \bibinfo {pages} {14118} (\bibinfo {year}
  {2018})}\BibitemShut {NoStop}%
\bibitem [{\citenamefont {Joos}\ \emph {et~al.}(2013)\citenamefont {Joos},
  \citenamefont {Zeh}, \citenamefont {Kiefer}, \citenamefont {Giulini},
  \citenamefont {Kupsch},\ and\ \citenamefont {Stamatescu}}]{c13}%
  \BibitemOpen
  \bibfield  {author} {\bibinfo {author} {\bibfnamefont {E.}~\bibnamefont
  {Joos}}, \bibinfo {author} {\bibfnamefont {H.~D.}\ \bibnamefont {Zeh}},
  \bibinfo {author} {\bibfnamefont {C.}~\bibnamefont {Kiefer}}, \bibinfo
  {author} {\bibfnamefont {D.~J.}\ \bibnamefont {Giulini}}, \bibinfo {author}
  {\bibfnamefont {J.}~\bibnamefont {Kupsch}},\ and\ \bibinfo {author}
  {\bibfnamefont {I.-O.}\ \bibnamefont {Stamatescu}},\ }\href@noop {} {\emph
  {\bibinfo {title} {Decoherence and the Appearance of a Classical World in
  Quantum Theory}}}\ (\bibinfo  {publisher} {Springer, Berlin},\ \bibinfo
  {year} {2013})\BibitemShut {NoStop}%
\bibitem [{\citenamefont {Montakhab}\ and\ \citenamefont
  {Asadian}(2008)}]{c14}%
  \BibitemOpen
  \bibfield  {author} {\bibinfo {author} {\bibfnamefont {A.}~\bibnamefont
  {Montakhab}}\ and\ \bibinfo {author} {\bibfnamefont {A.}~\bibnamefont
  {Asadian}},\ }\bibfield  {title} {\bibinfo {title} {Dynamics of global
  entanglement under decoherence},\ }\href
  {https://doi.org/10.1103/PhysRevA.77.062322} {\bibfield  {journal} {\bibinfo
  {journal} {Phys. Rev. A}\ }\textbf {\bibinfo {volume} {77}},\ \bibinfo
  {pages} {062322} (\bibinfo {year} {2008})}\BibitemShut {NoStop}%
\bibitem [{\citenamefont {Masanes}(2006)}]{c15}%
  \BibitemOpen
  \bibfield  {author} {\bibinfo {author} {\bibfnamefont {L.}~\bibnamefont
  {Masanes}},\ }\bibfield  {title} {\bibinfo {title} {All bipartite entangled
  states are useful for information processing},\ }\href
  {https://doi.org/10.1103/PhysRevLett.96.150501} {\bibfield  {journal}
  {\bibinfo  {journal} {Phys. Rev. Lett.}\ }\textbf {\bibinfo {volume} {96}},\
  \bibinfo {pages} {150501} (\bibinfo {year} {2006})}\BibitemShut {NoStop}%
\bibitem [{\citenamefont {Ezhov}\ and\ \citenamefont {Berman}(2001)}]{c16}%
  \BibitemOpen
  \bibfield  {author} {\bibinfo {author} {\bibfnamefont {A.~A.}\ \bibnamefont
  {Ezhov}}\ and\ \bibinfo {author} {\bibfnamefont {G.}~\bibnamefont {Berman}},\
  }\bibfield  {title} {\bibinfo {title} {{Role of interference and entanglement
  in quantum neural processing}},\ }in\ \href
  {https://doi.org/10.1117/12.449168} {\emph {\bibinfo {booktitle} {Electronics
  and Structures for MEMS II}}},\ Vol.\ \bibinfo {volume} {4591},\ \bibinfo
  {editor} {edited by\ \bibinfo {editor} {\bibfnamefont {N.~W.}\ \bibnamefont
  {Bergmann}}, \bibinfo {editor} {\bibfnamefont {D.}~\bibnamefont {Abbott}},
  \bibinfo {editor} {\bibfnamefont {A.}~\bibnamefont {Hariz}},\ and\ \bibinfo
  {editor} {\bibfnamefont {V.~K.}\ \bibnamefont {Varadan}}},\ \bibinfo
  {organization} {International Society for Optics and Photonics}\ (\bibinfo
  {publisher} {SPIE},\ \bibinfo {year} {2001})\ pp.\ \bibinfo {pages} {367 --
  379}\BibitemShut {NoStop}%
\bibitem [{\citenamefont {Band}\ and\ \citenamefont {Avishai}(2013)}]{c18}%
  \BibitemOpen
  \bibfield  {author} {\bibinfo {author} {\bibfnamefont {Y.~B.}\ \bibnamefont
  {Band}}\ and\ \bibinfo {author} {\bibfnamefont {Y.}~\bibnamefont {Avishai}},\
  }
  {\emph {\bibinfo {title} {Quantum Mechanics with Applications to
  Nanotechnology and Information Science}}}\ (\bibinfo  {publisher} {Academic
  Press},\ \bibinfo {address} {Amsterdam},\ \bibinfo {year} {2013})\BibitemShut
  {NoStop}%
\bibitem [{\citenamefont {Andrews}(2008)}]{c19}%
  \BibitemOpen
  \bibfield  {author} {\bibinfo {author} {\bibfnamefont {D.~L.}\ \bibnamefont
  {Andrews}},\ }\href
  {https://doi.org/https://doi.org/10.1016/B978-0-12-374027-4.50001-2} {\emph
  {\bibinfo {title} {Structured Light and Its Applications}}}\ (\bibinfo
  {publisher} {Academic Press},\ \bibinfo {address} {Burlington},\ \bibinfo
  {year} {2008})\BibitemShut {NoStop}%
\bibitem [{\citenamefont {Viola}\ and\ \citenamefont {Lloyd}(1998)}]{c20}%
  \BibitemOpen
  \bibfield  {author} {\bibinfo {author} {\bibfnamefont {L.}~\bibnamefont
  {Viola}}\ and\ \bibinfo {author} {\bibfnamefont {S.}~\bibnamefont {Lloyd}},\
  }\bibfield  {title} {\bibinfo {title} {Dynamical suppression of decoherence
  in two-state quantum systems},\ }\href
  {https://doi.org/10.1103/PhysRevA.58.2733} {\bibfield  {journal} {\bibinfo
  {journal} {Phys. Rev. A}\ }\textbf {\bibinfo {volume} {58}},\ \bibinfo
  {pages} {2733} (\bibinfo {year} {1998})}\BibitemShut {NoStop}%
\bibitem [{\citenamefont {Viola}\ \emph {et~al.}(1999)\citenamefont {Viola},
  \citenamefont {Knill},\ and\ \citenamefont {Lloyd}}]{c21}%
  \BibitemOpen
  \bibfield  {author} {\bibinfo {author} {\bibfnamefont {L.}~\bibnamefont
  {Viola}}, \bibinfo {author} {\bibfnamefont {E.}~\bibnamefont {Knill}},\ and\
  \bibinfo {author} {\bibfnamefont {S.}~\bibnamefont {Lloyd}},\ }\bibfield
  {title} {\bibinfo {title} {Dynamical decoupling of open quantum systems},\
  }\href {https://doi.org/10.1103/PhysRevLett.82.2417} {\bibfield  {journal}
  {\bibinfo  {journal} {Phys. Rev. Lett.}\ }\textbf {\bibinfo {volume} {82}},\
  \bibinfo {pages} {2417} (\bibinfo {year} {1999})}\BibitemShut {NoStop}%
\bibitem [{\citenamefont {Uhrig}(2007)}]{c22}%
  \BibitemOpen
  \bibfield  {author} {\bibinfo {author} {\bibfnamefont {G.~S.}\ \bibnamefont
  {Uhrig}},\ }\bibfield  {title} {\bibinfo {title} {Keeping a quantum bit alive
  by optimized $\ensuremath{\pi}$-pulse sequences},\ }\href
  {https://doi.org/10.1103/PhysRevLett.98.100504} {\bibfield  {journal}
  {\bibinfo  {journal} {Phys. Rev. Lett.}\ }\textbf {\bibinfo {volume} {98}},\
  \bibinfo {pages} {100504} (\bibinfo {year} {2007})}\BibitemShut {NoStop}%
\bibitem [{\citenamefont {Khodjasteh}\ \emph {et~al.}(2010)\citenamefont
  {Khodjasteh}, \citenamefont {Lidar},\ and\ \citenamefont {Viola}}]{c23}%
  \BibitemOpen
  \bibfield  {author} {\bibinfo {author} {\bibfnamefont {K.}~\bibnamefont
  {Khodjasteh}}, \bibinfo {author} {\bibfnamefont {D.~A.}\ \bibnamefont
  {Lidar}},\ and\ \bibinfo {author} {\bibfnamefont {L.}~\bibnamefont {Viola}},\
  }\bibfield  {title} {\bibinfo {title} {Arbitrarily accurate dynamical control
  in open quantum systems},\ }\href
  {https://doi.org/10.1103/PhysRevLett.104.090501} {\bibfield  {journal}
  {\bibinfo  {journal} {Phys. Rev. Lett.}\ }\textbf {\bibinfo {volume} {104}},\
  \bibinfo {pages} {090501} (\bibinfo {year} {2010})}\BibitemShut {NoStop}%
\bibitem [{\citenamefont {West}\ \emph {et~al.}(2010)\citenamefont {West},
  \citenamefont {Fong},\ and\ \citenamefont {Lidar}}]{c24}%
  \BibitemOpen
  \bibfield  {author} {\bibinfo {author} {\bibfnamefont {J.~R.}\ \bibnamefont
  {West}}, \bibinfo {author} {\bibfnamefont {B.~H.}\ \bibnamefont {Fong}},\
  and\ \bibinfo {author} {\bibfnamefont {D.~A.}\ \bibnamefont {Lidar}},\
  }\bibfield  {title} {\bibinfo {title} {Near-optimal dynamical decoupling of a
  qubit},\ }\href {https://doi.org/10.1103/PhysRevLett.104.130501} {\bibfield
  {journal} {\bibinfo  {journal} {Phys. Rev. Lett.}\ }\textbf {\bibinfo
  {volume} {104}},\ \bibinfo {pages} {130501} (\bibinfo {year}
  {2010})}\BibitemShut {NoStop}%
\bibitem [{\citenamefont {De~Lange}\ \emph {et~al.}(2010)\citenamefont
  {De~Lange}, \citenamefont {Wang}, \citenamefont {Riste}, \citenamefont
  {Dobrovitski},\ and\ \citenamefont {Hanson}}]{c25}%
  \BibitemOpen
  \bibfield  {author} {\bibinfo {author} {\bibfnamefont {G.}~\bibnamefont
  {De~Lange}}, \bibinfo {author} {\bibfnamefont {Z.}~\bibnamefont {Wang}},
  \bibinfo {author} {\bibfnamefont {D.}~\bibnamefont {Riste}}, \bibinfo
  {author} {\bibfnamefont {V.}~\bibnamefont {Dobrovitski}},\ and\ \bibinfo
  {author} {\bibfnamefont {R.}~\bibnamefont {Hanson}},\ }\bibfield  {title}
  {\bibinfo {title} {Universal dynamical decoupling of a single solid-state
  spin from a spin bath},\ }\href {https://doi.org/10.1126/science.1192739}
  {\bibfield  {journal} {\bibinfo  {journal} {Science}\ }\textbf {\bibinfo
  {volume} {330}},\ \bibinfo {pages} {60} (\bibinfo {year} {2010})}\BibitemShut
  {NoStop}%
\bibitem [{\citenamefont {Jiang}\ and\ \citenamefont {Imambekov}(2011)}]{c26}%
  \BibitemOpen
  \bibfield  {author} {\bibinfo {author} {\bibfnamefont {L.}~\bibnamefont
  {Jiang}}\ and\ \bibinfo {author} {\bibfnamefont {A.}~\bibnamefont
  {Imambekov}},\ }\bibfield  {title} {\bibinfo {title} {Universal dynamical
  decoupling of multiqubit states from environment},\ }\href
  {https://doi.org/10.1103/PhysRevA.84.060302} {\bibfield  {journal} {\bibinfo
  {journal} {Phys. Rev. A}\ }\textbf {\bibinfo {volume} {84}},\ \bibinfo
  {pages} {060302(R)} (\bibinfo {year} {2011})}\BibitemShut {NoStop}%
\bibitem [{\citenamefont {Wang}\ \emph {et~al.}(2011)\citenamefont {Wang},
  \citenamefont {Rong}, \citenamefont {Feng}, \citenamefont {Xu}, \citenamefont
  {Chong}, \citenamefont {Su}, \citenamefont {Gong},\ and\ \citenamefont
  {Du}}]{c27}%
  \BibitemOpen
  \bibfield  {author} {\bibinfo {author} {\bibfnamefont {Y.}~\bibnamefont
  {Wang}}, \bibinfo {author} {\bibfnamefont {X.}~\bibnamefont {Rong}}, \bibinfo
  {author} {\bibfnamefont {P.}~\bibnamefont {Feng}}, \bibinfo {author}
  {\bibfnamefont {W.}~\bibnamefont {Xu}}, \bibinfo {author} {\bibfnamefont
  {B.}~\bibnamefont {Chong}}, \bibinfo {author} {\bibfnamefont {J.-H.}\
  \bibnamefont {Su}}, \bibinfo {author} {\bibfnamefont {J.}~\bibnamefont
  {Gong}},\ and\ \bibinfo {author} {\bibfnamefont {J.}~\bibnamefont {Du}},\
  }\bibfield  {title} {\bibinfo {title} {Preservation of bipartite
  pseudoentanglement in solids using dynamical decoupling},\ }\href
  {https://doi.org/10.1103/PhysRevLett.106.040501} {\bibfield  {journal}
  {\bibinfo  {journal} {Phys. Rev. Lett.}\ }\textbf {\bibinfo {volume} {106}},\
  \bibinfo {pages} {040501} (\bibinfo {year} {2011})}\BibitemShut {NoStop}%
\bibitem [{\citenamefont {Chaudhry}(2014)}]{c28}%
  \BibitemOpen
  \bibfield  {author} {\bibinfo {author} {\bibfnamefont {A.~Z.}\ \bibnamefont
  {Chaudhry}},\ }\bibfield  {title} {\bibinfo {title} {Utilizing
  nitrogen-vacancy centers to measure oscillating magnetic fields},\ }\href
  {https://doi.org/10.1103/PhysRevA.90.042104} {\bibfield  {journal} {\bibinfo
  {journal} {Phys. Rev. A}\ }\textbf {\bibinfo {volume} {90}},\ \bibinfo
  {pages} {042104} (\bibinfo {year} {2014})}\BibitemShut {NoStop}%
\bibitem [{\citenamefont {Manovitz}\ \emph {et~al.}(2017)\citenamefont
  {Manovitz}, \citenamefont {Rotem}, \citenamefont {Shaniv}, \citenamefont
  {Cohen}, \citenamefont {Shapira}, \citenamefont {Akerman}, \citenamefont
  {Retzker},\ and\ \citenamefont {Ozeri}}]{c29}%
  \BibitemOpen
  \bibfield  {author} {\bibinfo {author} {\bibfnamefont {T.}~\bibnamefont
  {Manovitz}}, \bibinfo {author} {\bibfnamefont {A.}~\bibnamefont {Rotem}},
  \bibinfo {author} {\bibfnamefont {R.}~\bibnamefont {Shaniv}}, \bibinfo
  {author} {\bibfnamefont {I.}~\bibnamefont {Cohen}}, \bibinfo {author}
  {\bibfnamefont {Y.}~\bibnamefont {Shapira}}, \bibinfo {author} {\bibfnamefont
  {N.}~\bibnamefont {Akerman}}, \bibinfo {author} {\bibfnamefont
  {A.}~\bibnamefont {Retzker}},\ and\ \bibinfo {author} {\bibfnamefont
  {R.}~\bibnamefont {Ozeri}},\ }\bibfield  {title} {\bibinfo {title} {Fast
  dynamical decoupling of the m\o{}lmer-s\o{}rensen entangling gate},\ }\href
  {https://doi.org/10.1103/PhysRevLett.119.220505} {\bibfield  {journal}
  {\bibinfo  {journal} {Phys. Rev. Lett.}\ }\textbf {\bibinfo {volume} {119}},\
  \bibinfo {pages} {220505} (\bibinfo {year} {2017})}\BibitemShut {NoStop}%
\bibitem [{\citenamefont {Pokharel}\ \emph {et~al.}(2018)\citenamefont
  {Pokharel}, \citenamefont {Anand}, \citenamefont {Fortman},\ and\
  \citenamefont {Lidar}}]{c30}%
  \BibitemOpen
  \bibfield  {author} {\bibinfo {author} {\bibfnamefont {B.}~\bibnamefont
  {Pokharel}}, \bibinfo {author} {\bibfnamefont {N.}~\bibnamefont {Anand}},
  \bibinfo {author} {\bibfnamefont {B.}~\bibnamefont {Fortman}},\ and\ \bibinfo
  {author} {\bibfnamefont {D.~A.}\ \bibnamefont {Lidar}},\ }\bibfield  {title}
  {\bibinfo {title} {Demonstration of fidelity improvement using dynamical
  decoupling with superconducting qubits},\ }\href
  {https://doi.org/10.1103/PhysRevLett.121.220502} {\bibfield  {journal}
  {\bibinfo  {journal} {Phys. Rev. Lett.}\ }\textbf {\bibinfo {volume} {121}},\
  \bibinfo {pages} {220502} (\bibinfo {year} {2018})}\BibitemShut {NoStop}%
\bibitem [{\citenamefont {Chaudhry}\ and\ \citenamefont {Gong}(2013)}]{c32}%
  \BibitemOpen
  \bibfield  {author} {\bibinfo {author} {\bibfnamefont {A.~Z.}\ \bibnamefont
  {Chaudhry}}\ and\ \bibinfo {author} {\bibfnamefont {J.}~\bibnamefont
  {Gong}},\ }\bibfield  {title} {\bibinfo {title} {Amplification and
  suppression of system-bath-correlation effects in an open many-body system},\
  }\href {https://doi.org/10.1103/PhysRevA.87.012129} {\bibfield  {journal}
  {\bibinfo  {journal} {Phys. Rev. A}\ }\textbf {\bibinfo {volume} {87}},\
  \bibinfo {pages} {012129} (\bibinfo {year} {2013})}\BibitemShut {NoStop}%
\bibitem [{\citenamefont {Fanchini}\ \emph {et~al.}(2007)\citenamefont
  {Fanchini}, \citenamefont {Hornos},\ and\ \citenamefont {Napolitano}}]{c31}%
  \BibitemOpen
  \bibfield  {author} {\bibinfo {author} {\bibfnamefont {F.~F.}\ \bibnamefont
  {Fanchini}}, \bibinfo {author} {\bibfnamefont {J.~E.~M.}\ \bibnamefont
  {Hornos}},\ and\ \bibinfo {author} {\bibfnamefont {R{.}~d.~J.}\ \bibnamefont
  {Napolitano}},\ }\bibfield  {title} {\bibinfo {title} {Continuously
  decoupling single-qubit operations from a perturbing thermal bath of scalar
  bosons},\ }\href {https://doi.org/10.1103/PhysRevA.75.022329} {\bibfield
  {journal} {\bibinfo  {journal} {Phys. Rev. A}\ }\textbf {\bibinfo {volume}
  {75}},\ \bibinfo {pages} {022329} (\bibinfo {year} {2007})}\BibitemShut
  {NoStop}%
\bibitem [{\citenamefont {Chaudhry}\ and\ \citenamefont {Gong}(2012)}]{c33}%
  \BibitemOpen
  \bibfield  {author} {\bibinfo {author} {\bibfnamefont {A.~Z.}\ \bibnamefont
  {Chaudhry}}\ and\ \bibinfo {author} {\bibfnamefont {J.}~\bibnamefont
  {Gong}},\ }\bibfield  {title} {\bibinfo {title} {Decoherence control:
  Universal protection of two-qubit states and two-qubit gates using continuous
  driving fields},\ }\href {https://doi.org/10.1103/PhysRevA.85.012315}
  {\bibfield  {journal} {\bibinfo  {journal} {Phys. Rev. A}\ }\textbf {\bibinfo
  {volume} {85}},\ \bibinfo {pages} {012315} (\bibinfo {year}
  {2012})}\BibitemShut {NoStop}%
\bibitem [{\citenamefont {Fanchini}\ \emph {et~al.}(2015)\citenamefont
  {Fanchini}, \citenamefont {Napolitano}, \citenamefont
  {\ifmmode~\mbox{\c{C}}\else \c{C}\fi{}akmak},\ and\ \citenamefont
  {Caldeira}}]{c34}%
  \BibitemOpen
  \bibfield  {author} {\bibinfo {author} {\bibfnamefont {F.~F.}\ \bibnamefont
  {Fanchini}}, \bibinfo {author} {\bibfnamefont {R{.}~d.~J.}\ \bibnamefont
  {Napolitano}}, \bibinfo {author} {\bibfnamefont {B.}~\bibnamefont
  {\ifmmode~\mbox{\c{C}}\else \c{C}\fi{}akmak}},\ and\ \bibinfo {author}
  {\bibfnamefont {A.~O.}\ \bibnamefont {Caldeira}},\ }\bibfield  {title}
  {\bibinfo {title} {Protecting the $\sqrt{\mathit{swap}}$ operation from
  general and residual errors by continuous dynamical decoupling},\ }\href
  {https://doi.org/10.1103/PhysRevA.91.042325} {\bibfield  {journal} {\bibinfo
  {journal} {Phys. Rev. A}\ }\textbf {\bibinfo {volume} {91}},\ \bibinfo
  {pages} {042325} (\bibinfo {year} {2015})}\BibitemShut {NoStop}%
\bibitem [{\citenamefont {Bhaktavatsala~Rao}\ and\ \citenamefont
  {Kurizki}(2011)}]{KurizkiPRA2011}%
  \BibitemOpen
  \bibfield  {author} {\bibinfo {author} {\bibfnamefont {D.~D.}\ \bibnamefont
  {Bhaktavatsala~Rao}}\ and\ \bibinfo {author} {\bibfnamefont {G.}~\bibnamefont
  {Kurizki}},\ }\bibfield  {title} {\bibinfo {title} {From zeno to anti-zeno
  regime: Decoherence-control dependence on the quantum statistics of the
  bath},\ }\href {https://doi.org/10.1103/PhysRevA.83.032105} {\bibfield
  {journal} {\bibinfo  {journal} {Phys. Rev. A}\ }\textbf {\bibinfo {volume}
  {83}},\ \bibinfo {pages} {032105} (\bibinfo {year} {2011})}\BibitemShut
  {NoStop}%
\bibitem [{\citenamefont {Wootters}(1998)}]{c40}%
  \BibitemOpen
  \bibfield  {author} {\bibinfo {author} {\bibfnamefont {W.~K.}\ \bibnamefont
  {Wootters}},\ }\bibfield  {title} {\bibinfo {title} {Entanglement of
  formation of an arbitrary state of two qubits},\ }\href
  {https://doi.org/10.1103/PhysRevLett.80.2245} {\bibfield  {journal} {\bibinfo
   {journal} {Phys. Rev. Lett.}\ }\textbf {\bibinfo {volume} {80}},\ \bibinfo
  {pages} {2245} (\bibinfo {year} {1998})}\BibitemShut {NoStop}%
\end{thebibliography}
\end{document}